\documentclass[12pt,preprint]{aastex}

\slugcomment{accepted for publication in AJ (September 2003), First submission: August 2002}

\shorttitle{HST-WFPC2 Observations of Ultra-cool Dwarfs : A search for Companions}
\shortauthors{Bouy et al.}

\begin{document}

\title{Multiplicity of Nearby Free-floating Ultra-cool Dwarfs:\\ a HST-WFPC2 search for companions}

\author{Herv\'e Bouy}
\affil{E.S.O, Karl Schwarzschildstra\ss e 2, D-85748 Garching, Germany}
\email{hbouy@eso.org }

\author{Wolfgang Brandner}
\affil{Max-Planck Institut f\"ur Astronomie, K\"onigstuhl 17, D-69117 Heidelberg, Germany}
\email{brandner@mpia.de}

\author{Eduardo L. Mart\'\i n}
\affil{Institute for Astronomy, University of Hawai`i, 2680 Woodlawn Drive, Honolulu, HI 96822, USA}
\email{ege@teide.ifa.hawaii.edu}

\author{Xavier Delfosse}
\affil{Laboratoire d'Astrophysique de l'Observatoire de Grenoble, 414 rue de la piscine, F-38400 Saint Martin d'H\`ere, France}
\email{Xavier.Delfosse@obs.ujf-grenoble.fr}

\author{France Allard}
\affil{Centre de Recherche Astronomique de Lyon (UML 5574), Ecole Normale Sup\'erieure, 69364 Lyon Cedex 07, France}
\email{fallard@ens-lyon.fr}

\author{Gibor Basri}
\affil{University of California at Berkeley, Astronomy Department, MC 3411 Berkeley, CA 94720, USA}
\email{basri@astro.berkeley.edu}



\begin{abstract}
We present HST/WFPC2 observations of a sample of 134 
ultra-cool objects (spectral types later than M7) coming from the DENIS, 2MASS and SDSS surveys, 
with distances estimated to range from 7 pc to 105 pc. Fifteen new ultra-cool binary candidates are reported here. Eleven known binaries are confirmed and orbital motion is detected in some of them. We estimate that the closest binary systems in this sample have periods between 5 and 20 years, and thus dynamical masses will be derived in the near future. For the calculation of binary frequency we restrict ourselves to systems with distances less than 20 pc. After correction of the binaries bias, we find a ratio of visual binaries (at the HST limit of detection) of around 10\%, and that $\sim$15\% of the 26 objects within 20 parsecs are binary systems with separations between 1 and 8 A.U. The observed frequency of ultra-cool binaries is similar than that of binaries with G-type primaries in the separation range from 2.1 A.U. to 140 A.U. There is also a clear deficit of ultra-cool binaries with separations greater than 15 A.U., and  a possible tendency for the  binaries to have mass ratios near unity. Most systems have indeed visual and near-infrared brightness ratios between 1 and 0.3. We discuss our results in the framework of current scenarios for the formation and evolution of free-floating brown dwarfs.
\end{abstract}

\keywords{stars: very low mass,  ultra-cool dwarfs, brown dwarfs, binary}

\section{Introduction}

Effective strategies to detect brown dwarfs are proper motion surveys (e.g \citet{ruiz97}), wide field CCD surveys in star forming regions, open clusters like the Pleiades (e.g \citet{stauffer94, zapa97, bouvier98}), or the new generation of optical and near-infrared all sky surveys like the Sloan Digital Sky Survey \citep{york00}, the DEep Near Infrared Survey \citep{denis}, and the 2\,Micron All Sky Survey \citep{2mass}. So far, DENIS has produced a list of $\sim$ 300 nearby very low-mass objects \citep{delfosse97, delfosse99, xavier2002, martin99a}, and a similar number has been detected by 2MASS \citep{kirk97, kirk99, kirk01, burgasser99, gizis00, kirk02}. To date, several hundreds of nearby free-floating ultra-cool and brown dwarfs have been identified, and many more candidates are waiting for confirmation. This nearby sample is ideal for resolving ultra-cool and brown dwarfs binaries, and is large enough for statistical studies.

While systematic surveys of their physical properties are just starting, we already have indications that binary brown dwarfs are not rare. \objectname[PPL15]{PPL 15}, the first brown dwarf in the Pleiades with confirmed lithium \citep{basri96}, turned out to be a spectroscopic binary \citep{basrimartin99}. Several direct imaging surveys in the field or in open clusters using the high spatial resolution provided by both HST and/or adaptive optics led to the discovery of a significant number of binaries \citep{martin97,martin99a,martin2000, koerner99,reid01, neuhauser02, close02a, close02b, potter02,burgasser03, gizis03, close03}.

There are several key questions which need to be answered and for which brown dwarf binaries are an important piece of the puzzle: where do free-floating brown dwarfs originate? Are they ejected stellar embryos \citep{reipurth01,bate02,delgado03}, or do they form more isolated like ordinary stars due to fragmentation of collapsing molecular cloud clumps \citep{bodenheimer99} ?

One can then ask, how the binary properties of brown dwarfs, such as the frequency, distribution of separations, distribution of mass-ratios (is there a lower mass limit for companions to brown dwarfs ?), or the relation between orbital period and eccentricity, compare to the properties of stellar binaries. These properties hold important clues on the origin of (binary) brown dwarfs. An agreement of brown dwarf and stellar binary properties would suggest the same formation mechanism for both types of objects. 

Resolved brown dwarf binaries provide a very valuable opportunity to measure sub-stellar dynamical masses. The first such measurement has been reported by \citet{lane01} in the brown dwarf binary Gliese 569B, discovered by \citet{martin2000}. Dynamical masses for brown dwarfs of different ages are very much needed, as the mass estimates currently depend on untested theoretical evolutionary tracks, model atmospheres, and assumptions on the internal structure of brown dwarfs. In particular, a brown dwarf with a given age and mass might have very similar colours and luminosity as a younger brown dwarf with a lower mass. This degeneracy in the mass-luminosity relation for sub-stellar objects makes it very hard to pin down the physical properties of brown dwarfs, and to achieve the interplay between observation and theory which is necessary in order to improve, adapt, and fine-tune models, and to guide the interpretation of the observations. A search for binary brown dwarfs, and a detailed study of their properties directly addresses these questions.

The Pleiades, the first cluster in which a significant population of (coeval) brown dwarfs was identified, is at a distance of 125 pc. This makes it hard to detect and resolve binary brown dwarfs which separations less than 20 A.U. Orbital periods of resolved brown dwarf binaries in the Pleiades will be excessively long ($\ge$ several 100 yrs), making it impractical to compute their orbits and to derive dynamical masses. A recent search for binaries among 34 very low mass members of the Pleiades by \citet{martin2000} with HST-WFPC2 and Adaptive Optics system of CFHT did not lead to the discovery of any binary, indicating a lack of wide brown dwarfs binaries.

Nearby free-floating brown dwarfs do not form a coeval sample, but they are close enough to measure their distance (angular parallax) precisely, and - by using HST/WFPC2 - it is possible to detect and resolve binary brown dwarfs with separations down to 0.4 A.U.\ (0\farcs06 at 7\,pc).

The objects we study here have distances between 7 and 105 pc. For a prototypical brown dwarf binary with a separation of  0\farcs060, component masses of 0.045 M$_\odot$ and 0.030 M$_\odot$, and a circular orbit, the orbital period would be respectively between $\sim$ 1 and $\sim$ 100 years. Within the next 5 to 10 years dynamical masses for a wide range of brown dwarf binary companions could be determined.

In section 2, we describe the observations and the data analysis. 
In section 3 we present and discuss the results we obtained on each binary, 
in section 4 we perform an analysis of the results, and in section 5 we discuss these results in the framework of current models of formation and evolution of free-floating brown dwarfs.

\section{Observations and data analysis}

\subsection{Sample selection}
Our initial sample consists primarily of 34 objects detected by the near-infrared 
sky surveys DENIS, 2MASS and SDSS. They have been selected by analysing their positions in a colour-magnitude diagram, looking for the reddest objects. 

In order to increase the sample of objects and the quality of the statistical study, the total sample presented in this paper put together data from our own program (GO8720, P.I Brandner) with data from two other HST programs. We thus analyse data from the public HST archive coming from programs GO8146, P.I Reid, including 21 objects \citep{reid01}, and program GO8581, P.I Reid \citep{gizis03}, including 84 objects. The total sample thus had 134 objects.

The complete list of targets of the program is shown in Table \ref{targets}. Eleven of these objects were already previously identified as binary brown dwarfs.

\subsection{Observations}

The observations occurred between February 2000 and August 2001 during HST Cycles 8 and 9.
They have been carried out in snapshot mode. Each object was observed with the Planetary Camera of HST/WFPC2, in the F675W (600s) and F814W (300s) for our own program, in the F606W (100s) and F814W (300$+$350s) filters for program GO8146 (Cycle 8, P.I Reid, see \citet{reid01}) and in the F814W (100s,  200s or 400s) and F1042M (500s) filters for program GO8581 (Cycle 9, P.I Reid, \citet{gizis03}). Some objects have been observed twice : once during our own HST program and the second time during HST program GO9157 (P.I Mart\'\i n). Our targets are very red, thus the observations in F814W were sensitive to even lower mass companions than the observations in F675W, despite the shorter exposure time and the lower quantum efficiency of WFPC2 at longer wavelengths. For our program (GO8720), only one exposure was taken in each filter for each object, thus not allowing to reject cosmic rays automatically. This choice was done to minimise overheads and maximise exposure times.

\subsection{Data Analysis \label{dataanalysis}}

\subsubsection{Method}

We processed the data in two steps. We first identified the multiple systems either directly when resolved or by measuring the ellipticity of all the targets to look for elongated objects, using the IRAF \footnote{IRAF is distributed by National Optical Astronomy Observatories, which is operated by the Association of Universities for Research in Astronomy, Inc., under contract with the National Science Foundation.}  \emph{phot} task. 
All the objects with an ellipticity greater than 0.10 were suspected to be a binary, but all images have been inspected manually anyway to be sure we did not miss any candidate.

\begin{figure}
\epsscale{0.9}
\plotone{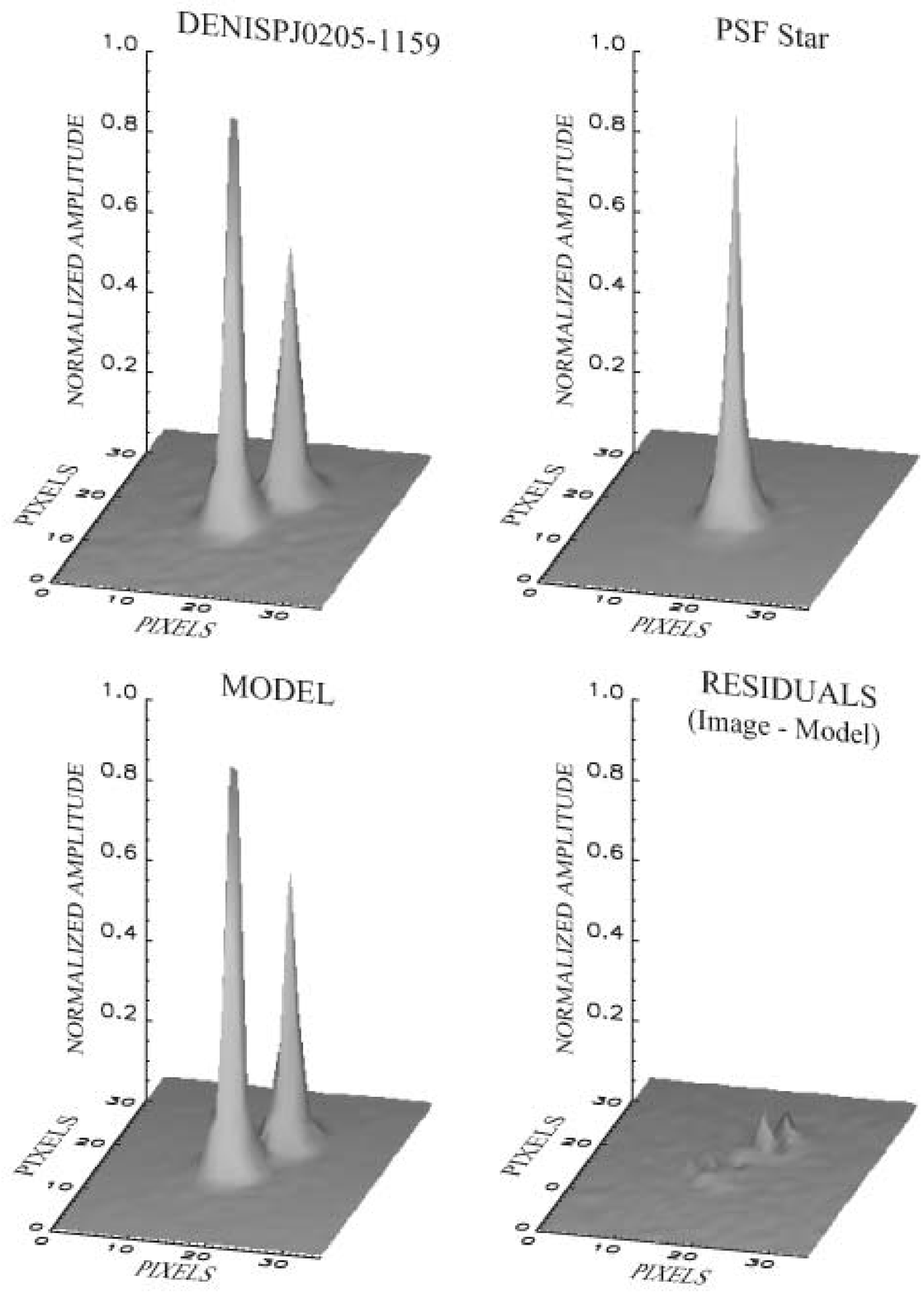}
\caption{Surface plots of the PSF fitting. Top left panel : Binary system DENISPJ0205-1159 with WFPC2/PC. Top right panel : PSF star from the field used to modelize the binary system. Bottom left panel : Model of DENISPJ0205-1159 built with the PSF Star. Bottom right panel : Residuals after subtraction of the model - All amplitudes are normalized, and sky has been subtracted. The images were obtained in the F814W filter. \label{fig1}}
\end{figure}

We used a custom-made PSF fitting program to compute the precise separation, 
position angle and flux ratio of each binary system. 
This program makes a non-linear fit of the binary system, fitting both components at the same time. 
It uses different PSF stars from throughout the sample and for each of them compute a model of the binary system (see Figure \ref{fig1}). 
The five free parameters for this model are the flux ratio and the pixel coordinates of the two components of the binary system. A cross correlation between the model and the binary system gives us the best values for these five parameters. To minimise the effects due to the slight position sensitivity of the PSF shape in the detector and the slight change of HST focus from one orbit to another, we built a library of 8 different PSF stars in each filter. As 8 PSF stars were not available in every field of view, we used the same library of PSF stars to analyse all the pictures. Comparison with analysis of images where 8 PSF stars were present in the field showed that this has no significant effect on the accuracy of the results.

\subsubsection{Calibration of the method}

To evaluate the accuracy of our PSF fitting program, and look for systematic errors, we used the program on 4950 simulated binaries, covering a range of 11 various input flux ratios (varying from 0.05 to 1.0, expressed as $F_{sec}/F_{prim}$),  135 various separations (varying from 0\farcs060 to 0\farcs600, by steps of 0\farcs004, thus oversampling the pixel scale) and 3 position angles (0, 22.5 and 45 degrees, measured in the detector's referential). It is important to note that these 3 angles, modulo 45 degrees, are sufficient to characterise 16 different position angles by symmetry in the detector. The simulated binaries were built using unresolved objects from the sample (different than the PSF stars used to make the PSF fitting). We made all these calibration in the F814W filter and expect it to be the same in the other filters. The S/N does not influence significantly the conclusions of the calibrations since the library of PSF stars used to make the PSF fitting spans a large range of S/N in each filters.

\begin{figure}
\epsscale{0.7}
\plotone{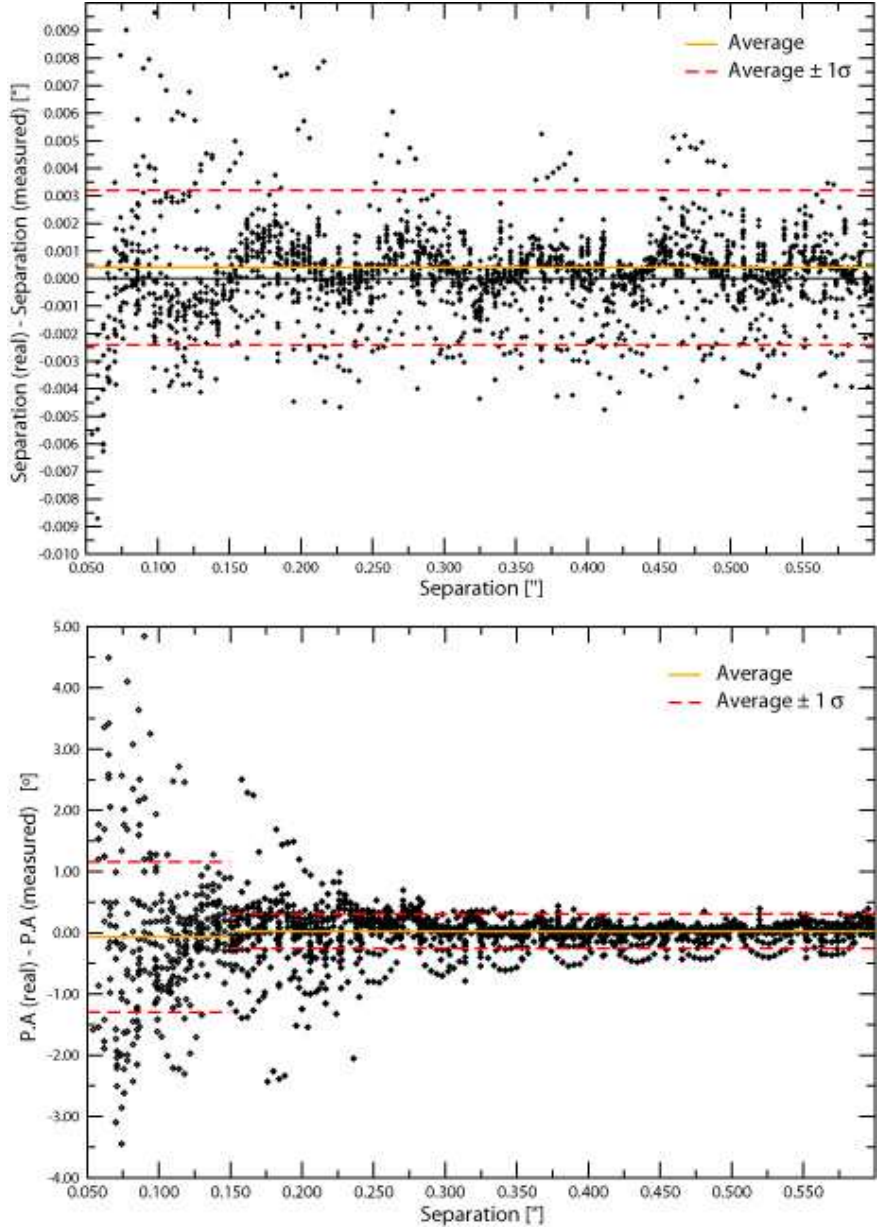}
\caption{Calibration of the systematic errors and uncertainties of the PSF fitting method. These figures shows the results we obtained using our PSF fitting method on simulated binaries with various flux ratios, separations and position angles. Top Panel: error on the separation (i.e difference between the real separation and the separation we measured) vs the real separation. Bottom Panel: same as top panel but for the error on the position angle. In both case the average and the standard deviation are represented. The systematic errors (corresponding to the average values) are almost null in both cases and we consider them negligible. The 1$\sigma$ uncertainty on the separation is 0\farcs0028. For the position angle, we divided the plot in two parts: before and after 0\farcs150. Before 0\farcs150, the  1$\sigma$ uncertainty is 1.2 degree. After 0\farcs150 it becomes only 0.3 degree. The periodic pattern has a period equal to the pixel scale (see section \ref{dataanalysis} for more details).\label{calib2}}
\end{figure}

Figure \ref{calib2} gives an overview of these calibrations for the separation and the position angle. Both plots show a periodic pattern which period is correlated to the pixel scale (P$\propto$2$\times$pixel scale, with a trigonometric scaling factor depending on the position angle). This effect is mostly due to the periodic errors introduced by the shift by a fraction of a pixel that we use to fit the binaries. 

It appears that the program gives good astrometric results. For our study, we will consider that the systematic errors on the separation and on the position angle are equal to the average of the errors, and that the 1$\sigma$ uncertainties are equal to the standard deviation of the errors. The systematic errors are negligible (less than 0\farcs0005 on the separation and almost zero on the position angle). The 1$\sigma$ uncertainty on the separation is 0\farcs0028. For the position angle it appears more appropriated to distinguish two cases: before and after 0\farcs150. The spreading of the values is indeed much larger in the first case than in the other (see Figure \ref{calib2}). This value of 0\farcs150 is certainly related to the size of the FWHM at these wavelengths ($\sim$ 2.4 pixels = 0\farcs110). The 1$\sigma$ uncertainty on the position angle  before 0\farcs150 is 1.2 degree, and becomes only 0.3 degree after 0\farcs150. As we could expect it (for symmetry reasons), there is no perceptible dependency on the position angle (except a slight change in the amplitude and period of the periodic pattern). The main variations are related to the difference of magnitude and of course to the separation.

\begin{figure}
\plotone{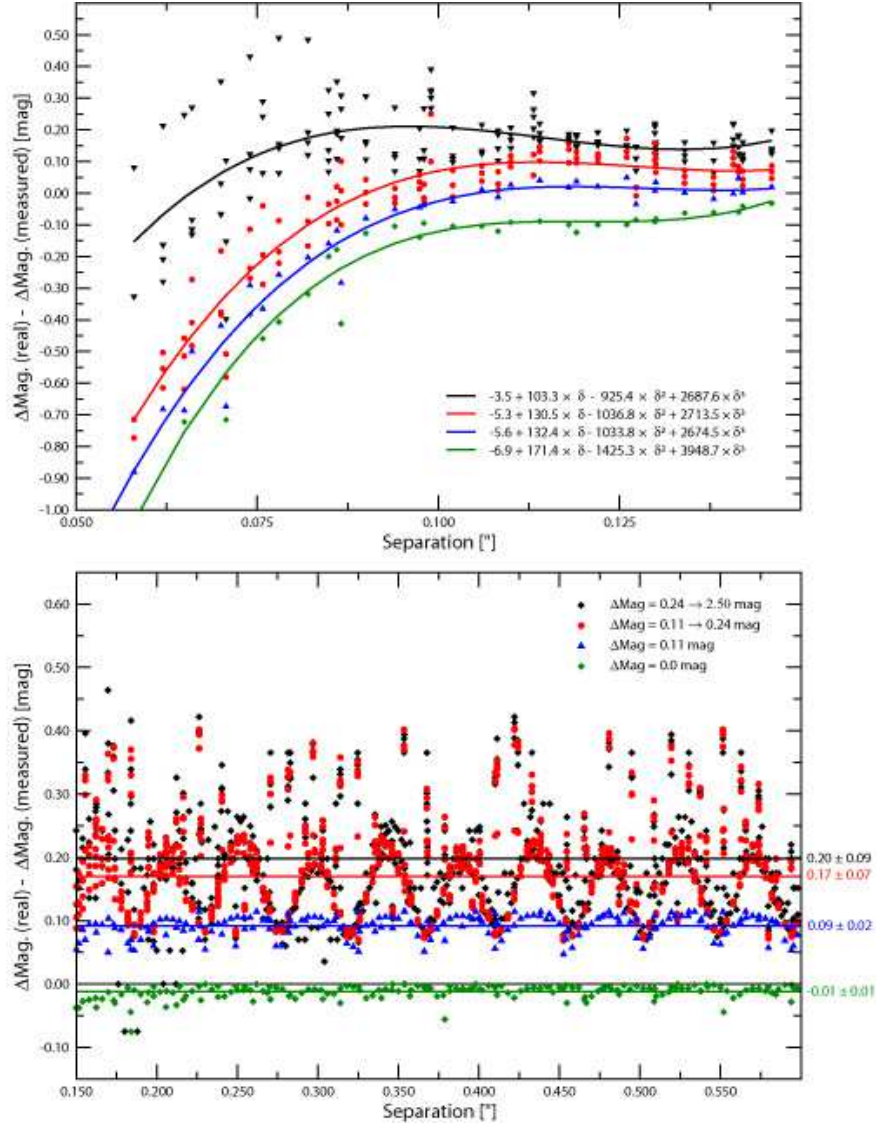}
\caption{Calibration of the systematic errors and uncertainties of the PSF fitting method. These figures shows the results we obtained using our PSF fitting method on simulated binaries with various flux ratios, separations and position angles. Top Panel: error on the difference of magnitude (i.e difference between the real difference of magnitude and the one measured) as a function of separation between 0\farcs060 and 0\farcs150. Bottom panel: same as top panel for the separation range between 0\farcs150 and 0\farcs600. \label{calib3}}

\end{figure}

The systematic errors and uncertainties on the difference of magnitudes require a more detailed analysis and description. Figure \ref{calib3} gives an overview of these results. The errors are very dependent on the difference of magnitude itself and of course on the separation, but are almost independent on the position angle (only the period and amplitude of the periodic pattern depends slightly on the position angle, see above).

Again it is more appropriate to distinguish two different parts in the range of separation: before and after 0\farcs150. The plots drawn in Figure \ref{calib3} show also clearly that it is necessary to distinguish four different cases in the range of differences of magnitudes: $\Delta$Mag=0.00 (hereafter case 1), $\Delta$Mag=0.11 (case 2), 0.11$< \Delta$Mag$\le$0.24 (case 3) and 0.24$< \Delta$Mag$\le$2.50 (case 4).

Cases 1 and 2 are the easiest to describe. The results are excellent after 0\farcs150: the systematic errors and 1$\sigma$ uncertainties are -0.01$\pm$0.01 mag for the case 1 and 0.09$\pm$0.02 mag for the case 2. Before 0\farcs150 the errors can be precisely described by a 3rd order law with dispersions of only 0.05 mag in both cases. We will not linger over a more detailed description of these two cases since they do not describe any of our objects.

The third and fourth cases are more difficult to analyse.

In the range of separation before 0\farcs150, the systematic errors in the third case can be relatively precisely described by a 3rd order law with a dispersion of 0.07 mag. On the same range of separation, the fourth case is not reproduced as well as the other cases by a 3rd order law. Nevertheless, the dispersion is 0.11 mag which is still reasonable. For the rest of our study, we will assimilate the systematic errors on these ranges of separation and differences of magnitude to the values given by the corresponding third order laws, and the 1$\sigma$ uncertainties to the corresponding dispersions (see Table \ref{table_calib1}).

In the range of separation after 0\farcs150, the third case shows also an obvious periodic pattern due mostly to the shift by a fraction of a pixel and to the normalization. Although it is obvious, this pattern cannot be easily fitted by a sinusoidal function without underestimating the systematic error, mainly because of the many points that are far away from the pattern. We thus decide to assimilate the systematic error on these values to the average (0.17 mag) and the 1$\sigma$ uncertainties to the standard deviation ($\pm$0.07 mag).

The pattern still appears but less obviously in the fourth case where the spreading of the errors is much larger. Here also we assimilate the systematic error to the average of these values (0.20 mag) and the 1$\sigma$ uncertainties to their standard deviation ($\pm$0.09 mag).

Table \ref{table_calib1} and \ref{table_calib2} gives an overview of the conclusions of these calibrations. All the values given in the text and in the tables have been corrected for the systematic errors and the given uncertainties correspond to the 1$\sigma$ uncertainties we calculated as explained in this section, unless specified explicitly.

\subsubsection{Accuracy of the results}

\begin{figure}
\plotone{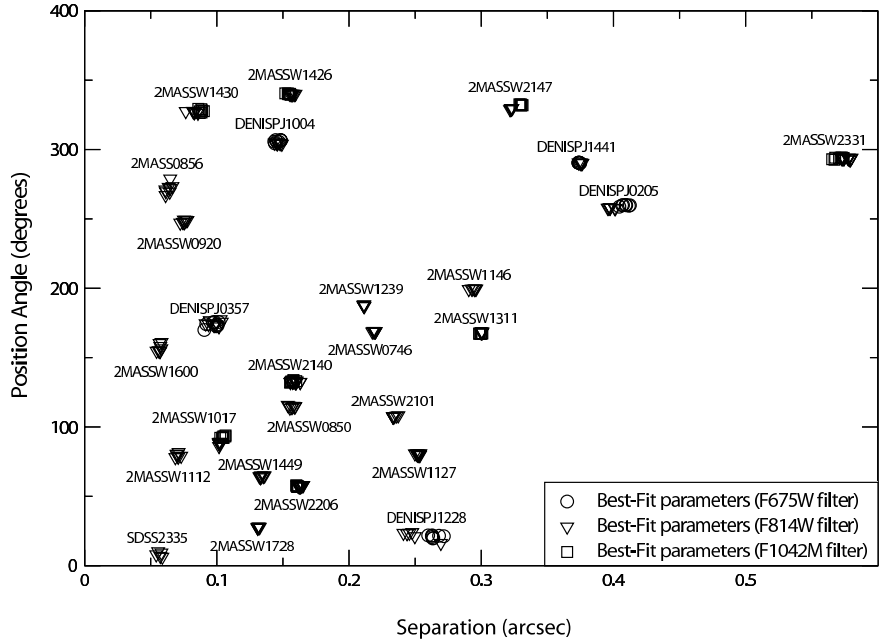}
\caption{Best Fit parameters (separation and position angle) obtained for each object using  different PSF stars for the fit. The eight values in each filters have been obtained using the library of eight different PSF stars. The parameters derived using the PSF fitting are in very good agreement for most of the objects, even at the smallest separations. The slight dispersion of the values for DENIS1228, DENIS0205 are explained respectively in paragraphs \ref{pd1228} and \ref{pd0205}. One can notice that the values in the F814W filter are usually less dispersed than in the F675W filter (worse S/N, worse sampling and coarser PSF). \label{pa_vs_sep}}
\end{figure}

An overview of the results is given in Table \ref{results} and Figure \ref{pa_vs_sep}, and examples are given in Figure \ref{fig1}. The uncertainties given in Table \ref{results} correspond to the 1$\sigma$ uncertainties calculated as explained in the previous section. As the observations in the F814W filter offer a much better S/N than in the F675W and F1042M filters, and a much better sampling than the F675W (the diffraction limit is indeed 0\farcs0703 in the F675W filter, and 0\farcs0860  in the F814W filter, and the pixel scale of the WFPC2/PC camera is 0\farcs0455), we consider them more accurate.
 
We therefore used only the values obtained with the F814W filter to compute the final parameters given in Table \ref{results}. For the data from the archive only the F814W and/or F1042M filters are available (the objects were too faint in the F606W filter).

The PSF at these wavelengths is under-sampled by the 0\farcs0455 pix$^{-1}$ plate scale of the Planetary Camera, thus not allowing to use deconvolution in the Fourier space. By using non-linear PSF fitting as described above, we were able to push the limit of detection down to $\sim$ 0\farcs060 arcsec for non-equal luminosity systems. Figure \ref{calib1} shows that we should have been able to detect all the binary systems with differences of magnitudes between $1.5 \lesssim \Delta m_{F814W} \lesssim 3.0$ mag easily even in the faintest cases, and binary systems with differences of magnitudes between  $3.0 \lesssim \Delta m_{F814W} \lesssim 5.5$ mag in the brightest cases.

In the next section we will comment on the results for each object.

\section{Results}

\subsection{Photometry, Spectral Classification and Distances \label{spt}}

An overview of values of photometry, spectral types, and distances for all the targets of the sample is given in Tables \ref{targets}, \ref{photom_go8720},\ref{photom_go8581} and \ref{photom_go8146}. The I,J,K$_{s}$ values of the DENIS objects come from the DENIS survey. The J,H,K$_{s}$ values of the 2MASS and some SDSS objects come from the 2MASS survey \citep{kirk00} and from 
Davy Kirkpatrick's on-line archive of L and T dwarfs \citep{kirk_url}. The I values of the 2MASS and SDSS targets have been calculated as follow : using the F675W and F814W magnitudes obtained in the data (see section \ref{colours}) and the I magnitudes of the DENIS objects of our program (see Table \ref{targets}), we derived the following relation: 
\begin{equation} \label{equation1}
I_{DENIS} = -0.81 + 1.02\times m_{F814W} + 0.28 \times (m_{F675W} - m_{F814W})
\end{equation}
with a dispersion of 0.18 mag (thus of the same order than the uncertainties on the DENIS magnitudes themselves) and used it to compute the I magnitudes\footnote{ I$_{DENIS}$ is very close to the I$_{Cousins}$ \citep{xavier_thesis}} given in Table \ref{targets}. 

The magnitudes have been measured with standard procedures using the aperture photometry task \emph{phot} in IRAF, using sky subtraction and a 3$\sigma$ rejection. We used a recommended aperture of 0\farcs5 in the case of unresolved objects, and an aperture of 1\farcs365 (30 pixels) in the case of binaries, to measure the total flux of the system. The counts were transformed to magnitudes using the relation : $m[mag]=-2.5 \cdot log(counts/exp)+ZP-0.1$ for the unresolved objects, where $counts$ is the number of counts measured with IRAF, $exp$ the exposure time, $ZP$ the Vega Zero Point magnitude ($ZP_{F675W}=22.042$ mag, $ZP_{F814W}=21.639$ mag, and $ZP_{F1042M}=16.148$ mag, \citet{hst_data_handbook}) and $-0.1$ is to correct from the finite to infinite aperture, according to the HST data handbook, and the relation: $m[mag]=-2.5 \cdot log(counts/exp)+ZP-0.028$ in the case of binaries, where $-0.028$ is the correction we evaluated as suggested in the HST data Handbook to correct from the finite to infinite aperture in that case. For multiple systems, we deduced the magnitude of each component using our values of flux ratios. In many cases we just had one image thus not allowing to remove cosmic ray events automatically. Nevertheless cosmic rays should not have influenced our photometry significantly since we would have been able to detect any cosmic ray event too close to the object to be removed by the 3$\sigma$ automatic rejection. This situation happened in only one case (2MASSW2147+1431, F1042M filter). All the cosmic ray events sufficiently far away from the object have been corrected by the 3$\sigma$ automatic rejection of the IRAF \emph{phot} task. The results are given in Tables \ref{results}, \ref{photom_go8720}, \ref{photom_go8581} and \ref{photom_go8146}. Uncertainties for the unresolved objects are 0.1 mag.

A few DENIS objects and most 2MASS and SDSS objects had accurate values of spectral types found in Davy Kirkpatrick's on-line archive of L and T dwarfs \citep{kirk_url}, and obtained through spectroscopic measurements. For the other objects, we used the \citet{dahn2002} spectral type vs. (I-J) colour relation to deduce the spectral types given in Table \ref{targets}. These latter spectral types will have to be confirmed by spectroscopic measurements. The sample thus covers a large range of spectral types, going from $\sim$M8 to $\sim$L8. 

Two efforts by the United States Naval Observatory and by an Australian group led by Chris Tinney are currently under way in order to derive precise angular parallaxes for these objects. Some of the targets already have published angular parallaxes and distances \citep{dahn2002}. For the unresolved objects without published trigonometric parallax, we evaluated the photometric parallax using the \citet{dahn2002} M$_{J}$ vs. Spectral Type relation when both J and spectral type where available, and the \citet{dahn2002} M$_{J}$ vs. (I-J) relation in the other cases. For the multiple systems, we evaluated the distance using the \citet{dahn2002} M$_{I}$ vs. Spectral Type relation and then multiplied the photometric distance obtained by a correction factor of $\sqrt{1+f}$, were $f$ is the flux ratio (in the F814W$\sim$I$_{C}$ filter), to correct for the bias introduced by the multiplicity. These results indicate that the ultra-cool dwarfs in the sample are at distances between 7\,pc and 105\,pc (cf. Table \ref{targets}). Some SDSS objects had no published values of spectral types and J,H,K$_{S}$ magnitudes, thus not allowing to evaluate their photometric distances. For these objects we assumed a typical distance of 20 pc.

\subsection{DENISPJ020529-115930 \label{pd0205}}
\objectname[DENISPJ020529-115930]{DENISPJ020529-115930} is a L7V brown dwarf in the Kirkpatrick classification scheme (L5 in the Mart\'\i n classification scheme). It has been discovered by \citet{delfosse97}, and announced as a binary by \citet{koerner99}, on the basis of K band observations at Keck (0\farcs51$\pm$0\farcs02 and 106$^{\circ} \pm$5$^{\circ}$ in July 1997, and 0\farcs51$\pm$0\farcs02 and 72$^{\circ} \pm$10$^{\circ}$ in January 1999). This object has also been observed in the 2MASS survey and is reported as \objectname[2MASSWJ0205293-115930]{2MASSWJ\-0205293\--115930}. \citet{leggett01} observed it in the IR with UKIRT on September 1999, and report a separation of 0\farcs35 $\pm$0\farcs03 and a position angle of 77$^{\circ} \pm$4$^{\circ}$.

\citet{delfosse97}, \citet{mclean01} and \citet{burgasser03} report the detection of methane absorption band in their IR spectra, which would imply a mass below the stellar limit, and an effective temperature below 1800K, as stated by \citet{schweitzer02}. \citet{basri2000} estimated this temperature to be $T_{eff}=1700 \sim 1800$ K. \citet{tokunaga99} also observed absorption feature in the spectra but attributed it to H$_{2}$ rather than CH$_{4}$.

An overview of interesting values currently known about \objectname[DENISPJ020529-115930]{DENISPJ020529-115930} is given in table \ref{table_d1228_d0205}.
On this particular object we had two sets of data, one coming from our HST/GO8720 program and one coming from the HST/GO9157 program.

The first set of data (GO8720) included two filter images. The agreement between the values obtained in each filter is not as good as for the other objects (cf. Figure \ref{pa_vs_sep}). We measured a separation of $409$ mas in the F675W filter and $398$ mas in the F814W filter. The discrepancy comes probably from the fact that the secondary is very elongated on both images. The object fell on a boundary between two pixels and its PSF is spread over two pixels, increasing the uncertainties. As the F814W image is more sensitive than the F675W one, and as the secondary is brighter (and thus easier to detect) in the F814W filter, we consider hereafter the corresponding value to be the best one. We measured differences of magnitudes of $\Delta m_{F675W}=1.47\pm0.19$ mag and $\Delta m_{F814W}=0.65\pm0.16$ mag, but these values could be altered by the fact that the PSF of the secondary is elongated.

The second set of data (GO9157) was better and more accurate values were obtained. Several exposures were available, thus allowing to reject cosmic rays. We measured a separation of  $370.0 \pm 2.8$ mas and a position angle of $255.8 \pm 0.3$ degrees. The difference of magnitude is  $\Delta m_{F814W}=0.63\pm0.09$ mag and is in good agreement with the previous one, suggesting that even if the PSF of the secondary was elongated, the value was not altered too much. Both \citet{leggett01, koerner99} report a difference of magnitude of about 0 mag in the IR. Figure \ref{fig1} shows clearly that this is not the case in the F814W filter. Our results on the separation of the system are in good agreement with the one given by \citet{leggett01} within the uncertainties but do not agree with the one given by \citet{koerner99}. We consider that our observations and the observations reported by \citet{leggett01}, based on higher spatial resolutions than the one of \citet{koerner99}, are more accurate.

To evaluate the orbital period of this system, we evaluate its semi-major axis by multiplying the projected separation by a statistical correction factor of 1.26 as explained in \citet{fischer92} (see Table \ref{table_periods}). Then assuming a distance of 19.8 pc, with a total mass of $\sim0.14$ M$_{\sun}$ (masses of the two components are unknown, but the presence of methane implies a mass less than 0.07 M$_{\sun}$ and the absence of lithium implies a mass greater than 0.055 M$_{\sun}$, assuming an age greater than 0.5 Gyr), and a semi-major axis of 9.2 A.U, the period of this system would be $\sim$ 74.6 years.

\subsection{DENISPJ035729.6-441730 \label{pd0357}}
\objectname[DENISPJ035729.6-441730]{DENISPJ035729.6-441730} is one of the new binary system candidates we have discovered in this survey. It consists of a very close binary, with a separation of $98 \pm 2.8$ mas and a position angle of $174.7 \pm 1.2$ degrees. The differences of magnitude are $\Delta m_{F675W}=1.23\pm0.11$ mag and $\Delta m_{F814W}=1.50\pm0.11$ mag, suggesting that the two components have slightly different masses. Despite the very small separation, (we are here very close to WFPC2/PC pixel scale), the values obtained for the separation and the position angle in the two filters are in very good agreement one with each other (cf. Figure \ref{pa_vs_sep}). The companion appears clearly after PSF fitting (see Figure \ref{close_bin}).

As described in section \ref{spt}, we evaluated its photometric distance corrected for multiplicity of 22.2 pc. To the projected separation of $22.2 \times 0.098 = 2.2$ A.U should correspond a semi-major axis of $2.2 \times 1.26 = 2.8$ A.U. Assuming a face-on orbit, with a total mass of $\sim0.20$ M$_{\sun}$ (masses are here unknown and we assume masses of about $\sim$ 0.1 M$_{\sun}$ for each component), and a semi-major axis of 2.8 A.U, the period of this system would be $\sim$ 10.5 years.

\subsection{2MASSW0746425+2000032}
\objectname[2MASSW0746425+2000032]{2MASSW0746425+2000032} (spectral type L0.5 \citep{kirk00}) has been discovered by \citet{kirk00}, and  was suggested to be a binary by \citet{reid00} based on its position in a colour-magnitude diagram. It has been confirmed as a multiple system by \citet{reid01}, with a separation of $0.22$ arcsec and a position angle of $15$ degrees. 
On the same data set we measured a separation of $219 \pm 2.8$ mas and a position angle of $168.8 \pm 0.3$ degrees (cf. Figure \ref{pa_vs_sep}). There is a discrepancy with the position angle given by \citet{reid01}. To measure the position angle, we used the positions of each component in the detector, and measured the corresponding position angle of $23.6$ degrees (in the detector). We then added $145.212$ degrees due to the orientation angle of the HST spacecraft (given by the \emph{ORIENTAT} header keyword), to obtain the $168.8$ degrees mentioned above. Figure \ref{all_bin} shows the image of 2MASSW0746425+2000032 and its orientation. Position angles are measured from the North to the East.

The distance of 2MASSW0746425+2000032 was estimated by trigonometric parallax measurements to be 12.3 pc. The projected radius of $219$ mas thus corresponds to a semi-major axis about $0.219 \times 12.3 \times 1.26 = 3.4$ A.U.

We measured $\Delta m_{F814W}=1.0\pm0.09$ mag. On the same set of data \citet{reid01} measured a difference of magnitude of $\Delta I=0.62$ mag, lower than the one we measure, but the uncertainties are not given.

Assuming a semi-major axis of 3.4 A.U, with a total mass of $\sim0.20$ M$_{\sun}$ (masses are here unknown and we assume masses of about $\sim$ 0.1 M$_{\sun}$ for each component), the period of this system would be $\sim$ 14.0 years.

\subsection{2MASSW0850359+105716}
\objectname[2MASSW0850359+105716]{2MASSW0850359+105716} was a known binary brown dwarf \citep{reid01} of spectral type L6 in the Kirkpatrick classification scheme (L5 in the Mart\'\i n classification scheme) implying $T_{eff} < 1800$ K. This object shows a strong lithium line \citep{kirk99} implying a mass $M \leq 0.06 M_{\sun}$. Its distance (25.6 pc) and proper motion ($\mu=144.7 \pm 2.0$ mas/yr) were estimated by USNO parallax measurements \citep{dahn2002}.
\citet{reid01} measured a separation of $0.16$ arcsec and a position angle of $250$ degrees on the 1st February 2001 using HST/WFPC2. On the same data set we measured a separation of $157 \pm 2.8$ mas and a position angle of $114.7 \pm 0.3$ degrees (cf. Figure \ref{pa_vs_sep}). There is again a discrepancy between the values measured for the position angle. Figure \ref{all_bin} shows the image of 2MASSW0850359+105716 and its orientation.

We measured difference of magnitude of $\Delta m_{F814W}=1.47\pm0.09$ mag. This implies again a slight difference in the masses of the two components. On the same set of data \citet{reid01} measured a difference of magnitude of $\Delta I=1.34$ mag, in good agreement with our value. Assuming a semi-major axis of $0.157 \times 41.0 \times 1.26 = 8.1$ A.U, with an assumed total mass of $\sim0.14$ M$_{\sun}$ (masses of the two components are unknown, but the presence of lithium in a L5 dwarf implies a mass less than 0.06 M$_{\sun}$ for one of them at least), the period of this system would be $\sim$ 61.6 years.

\subsection{2MASSW0856479+223518}
\objectname[2MASSW0856479+223518]{2MASSW0856479+223518} is a good candidate binary we report in this paper. It has been observed on the 24th of April 2001 in the F814W and F1042M filters, but was not resolved in the latter one. We measured a separation of $98 \pm 9$ mas and a position angle of $275 \pm 2.0$ degrees. The difference of magnitude is $\Delta m_{F814W}=2.76\pm0.11$ mag, suggesting that the two components have probably different masses. This large difference of magnitudes at such a small separation makes it very difficult to measure precisely the different parameters, explaining the higher uncertainties on the values. Nevertheless the two centroids appear clearly on the image (cf. Figure \ref{contour_plot}) and the PSF fitting makes appear a faint but obvious companion (see Figure \ref{close_bin}). As we reach here the limit of detection of our method, the multiplicity of this object should be considered with caution and further followup observations with higher spatial resolution are required in order to confirmed that it is a real binary.

As described in section \ref{spt}, we evaluated a photometric distance corrected for the multiplicity of $d=34.7$ pc, which leads to a semi-major axis of 4.4 A.U, and with a total mass of $\sim0.20$ M$_{\sun}$, to a period of $\sim$ 20.8 years (masses are here unknown and we assume masses of about $\sim$ 0.1 M$_{\sun}$ for each component).

\subsection{2MASSW0920122+351743}
\objectname[2MASSW0920122+351743]{2MASSW0920122+351743} (L6.5) has been identified as a brown dwarf by \citet{kirk00} and as a binary by \citet{reid01}. \citet{nakajima01} report the observation of methane in the H and K bands, which make more difficult the definition of the transition between L and T dwarfs.

For that object, \citet{reid01} report a separation of $75$ mas and a position angle of $90$ degrees on the 2nd of September 2000. On the same set of data we measured a separation of $75 \pm 2.8$ mas and a position angle of $248.5 \pm 1.2$ degrees (cf. Figure \ref{pa_vs_sep} and \ref{all_bin}). Once again there is a discrepancy in the position angle. Figure \ref{all_bin} shows the image of 2MASSW0920122+351743 and its orientation. We measured a difference of magnitude $\Delta m_{F814W}=0.88\pm0.11$ mag. On the same set of data \citet{reid01} measured a difference of magnitude of $\Delta I=0.44$ mag, lower than the one we measure, but the uncertainties are not given. Considering only our uncertainties, the two values are different by 4$\sigma$.

This makes of 2MASSW0920122+351743 one of the closest resolved binaries observed in the sample. Figure \ref{close_bin} shows the companion clearly appearing after PSF fitting. \citet{kirk00} estimated a distance of $21$ pc for the unresolved system. Using our photometric measurement we derive a distance of 16.7 pc. Correcting for multiplicity it gives a distance of $20.1$ pc for the multiple system. We can estimate the semi-major axis to be about $1.9$ A.U.

Assuming  a total mass of $\sim0.14$ M$_{\sun}$ (masses of the two components are unknown, but the presence of methane in this L6.5 dwarf implies a mass of less than 0.07 M$_{\sun}$), and a semi-major axis of 1.9 A.U, the period of this system would be $\sim$7.2 years.
 
\subsection{DENISPJ100428.3-114648 \label{d1004}}
\objectname[DENISPJ100428.3-114648]{DENISPJ100428.3-114648} is a new binary system candidate with a separation of $146 \pm 2.8 $ mas and position angle of $304.5 \pm 1.2$ degrees. It has also been reported in the 2MASS Survey as \objectname[2MASSW1004282-114648]{2MASSW1004282-114648}. The differences of magnitudes are $\Delta m_{F675W}=0.25\pm0.07$ mag and $\Delta m_{F814W}=0.66\pm0.11$ mag. Once again the agreement between the values for separation and position angle obtained in the two filters is very good (cf. Figure \ref{pa_vs_sep}). As explained in section \ref{spt}, we estimated its photometric distance corrected for binarity to be 46.8 pc.

Assuming a semi-major axis of 8.6 A.U, with a total mass of $\sim0.20$ M$_{\sun}$ (masses are here unknown and we assume masses of about $\sim$ 0.1 M$_{\sun}$ for each component), the period of this system would be $\sim$ 56.5 years.

\subsection{2MASSW1017075+130839}
\objectname[2MASSW1017075+130839]{2MASSW1017075+130839} is a new binary candidate we report in this paper. It has been observed on the 16th of April 2001 in the F1042M and F814W filters. We measure a separation of $104 \pm 2.8$ mas and a position angle of $92.6 \pm 1.2$ degrees. The difference of magnitude are $\Delta m_{F1042M}=0.74\pm0.11$ mag, and $\Delta m_{F814W}=0.77\pm0.11$ mag, suggesting that the two components have probably similar masses.

We evaluated the photometric distance corrected for binarity at $\sim$ 21.4 pc. Assuming a semi-major axis of 2.9 A.U, with a total mass of $\sim0.20$ M$_{\sun}$ (masses are here unknown and we assume masses of about $\sim$ 0.1 M$_{\sun}$ for each component), the period of this system would be $\sim$ 11.0 years.

\subsection{2MASSW1112257+354813}
\objectname[2MASSW1112257+354813]{2MASSW1112257+354813} is also one of the 15 new binary candidates we report in this paper. It has been observed on the 14th of February 2001 in the F814W and F1042M filters. For this object we measure a separation of $70 \pm 2.8$ mas and a position angle of $79.6 \pm 1.2$ degrees. The differences of magnitude are $\Delta m_{F814W}=1.07\pm0.11$ mag and $\Delta m_{F1042M}=1.04\pm0.11$ mag. Figure \ref{close_bin} shows the companion appearing clearly after PSF fitting.

This object is particularly interesting since 2MASSW1112257+354813 is also known as \objectname[Gl 417B]{Gl 417B}, confirmed as a L-dwarf and as companion of the G0-dwarf \objectname[Gl 417A]{Gl 417A} by \citet{kirk01}, who measured a separation between the G-dwarf primary and the 2MASSW1112257+354813 unresolved system of 90\farcs0 and a position angle of 245 degrees. For such a system to be stable, the ratio of the semi-major axis has to be greater than $\sim$5, which is here clearly the case ($\rho_{AB}/ \rho_{BC}=90/0.070 \sim 1300$). \citet{kirk01} estimated an age of 0.08-0.3 Gyr for the G-dwarf primary and, assuming that the primary and its companions are coeval, they computed a mass of 0.035$\pm$0.015$M_{\sun}$ for 2MASSW1112257+354813, well below the hydrogen burning limit. The G-dwarf primary appears clearly on the HST/WFPC2 image and saturates completely one of the four CCD of the HST/WFPC2 camera.

This triple system is thus similar to Gl 569Bab \citep{martin2000b, ken01} and HD130948 \citep{potter02} (see Table \ref{triple}).

The distance of the G-dwarf primary has been evaluated by trigonometric parallax ($\pi=$46.04 mas or d=21.7 pc; Hipparcos, \citet{perryman97}) and \citet{kirk00} assigned a spectral type of L4.5. Assuming a semi-major axis of $1.26 \times 21.7 \times 0.070 = 1.9$ A.U, with a total mass of $\sim0.20$ M$_{\sun}$ (masses are here unknown and we assume masses of about $\sim$ 0.1 M$_{\sun}$ for each component), the period of this system would be $\sim$ 5.8 years only. As its distance is known precisely, and as its age can be constrained by studying the G-dwarf primary, this system is a very good and very promising candidate to constrain theory and the mass/luminosity function. 

\subsection{2MASSW1127534+741107}
\objectname[2MASSW1127534+741107]{2MASSW1127534+741107} is a M8.0 dwarf \citep{gizis00} and also one of the 15 new binary candidates we report in this paper. It has been observed on the 5th of May 2001 in the F1042M and F814W filters. We measure a separation of $252.5 \pm 2.8$ mas and a position angle of $79.9 \pm 0.3$ degrees. The differences of magnitude between the primary and the secondary are $\Delta m_{F1042M}=0.39\pm0.07$ mag and $\Delta m_{F814W}=0.82\pm0.09$ mag, suggesting that the two components have probably similar masses.

We evaluated the photometric distance corrected for multiplicity at 14.6 pc. Assuming a semi-major axis of 4.8 A.U, with a total mass of $\sim0.20$ M$_{\sun}$ (masses are here unknown and we assume again masses of about $\sim$ 0.1 M$_{\sun}$ for each component), the period of this system would be $\sim$ 23.5 years.

\subsection{2MASSW1146344+223052}
\objectname[2MASSW1146344+223052]{2MASSW1146344+223052} has been discovered as a L3 dwarf by \citet{kirk99} and as a binary by \citet{koerner99}. It shows lithium lines and \citet{basri2000} estimated its effective temperature to $1950$ K. Its distance ($27.2$ pc) has been measured by trigonometric parallax \citep{dahn2002}. \citet{reid01} measured a separation of 0\farcs29 and a position angle of $199.5$ degrees. On the same set of data we measured a separation of $294 \pm 2.8 $ mas and a position angle of $199.5 \pm 0.3$ degrees (cf. Figure \ref{pa_vs_sep}). This time both position angle and separation are in agreement with the values given by \citet{reid01}. We also measured a difference of magnitude $\Delta m_{F814W}=0.75\pm0.09$ mag. On the same set of data \citet{reid01} measured a difference of magnitude of $\Delta I=0.31$ mag, again smaller than the one we measure, but the uncertainties are not given. Considering only our uncertainties, the two measurements are different by 5$\sigma$. Such a difference is much larger than the uncertainties on our measurement, as explained in section \ref{dataanalysis}, especially considering that with a separation of $\sim$0\farcs29 and a difference of magnitude less than 1.0, 2MASSW1146344+223052 was an ``easy'' case for the PSF fitting program.

Assuming a semi-major axis of 10.1 A.U, with a total mass of $\sim0.12$ M$_{\sun}$ (masses of the two components are unknown, but the presence of lithium in a L7 dwarf implies a mass equal or less than 0.06 M$_{\sun}$), the period of this system would be $\sim$ 92.6 years.

\subsection{DENISPJ122813.8-154711 \label{pd1228}}
\objectname[DENISPJ122813.8-154711]{DENISPJ122813.8-154711} has been discovered by \citet{delfosse97} and resolved for the first time by \citet{martin99b}. It is a L4.5 brown dwarf in the Mart\'\i n classification scheme and a L5 in the Kirkpatrick classification scheme \citep{kirk99}. This object has been observed and studied several times. We provide a summary of the astrometric and photometric measurements in Table \ref{table_d1228_d0205}.

This object has also been observed in the 2MASS survey and is reported as \objectname[2MASSWJ1228152-154734]{2MASSWJ\-1228152\--154734}. The designations are different mostly because the DENIS astrometric pipeline used at the time of discovery was not the final version and the uncertainties of the astrometry were high.

Lithium absorption has been reported by both \citet{martin97} and \citet{tinney97}, implying a mass M $\leq 0.06$ M$_{\sun}$. DENISPJ122813.8-154711 was the second field brown dwarf to be confirmed by the lithium test.

For this object again we had two sets of data. The results obtained with the first set are not very accurate because of coordinate mismatch. The target fell in a corner of one of the Wide Field Camera of WFPC2 instead of in the centre of the Planetary Camera (PC), where the distortions are more important and the pixel scale coarser, implying a worse sampling of the PSF and thus higher uncertainties. Thus the results in the two filters are not in very good agreement (cf. Table \ref{table_d1228_d0205} and Figure \ref{pa_vs_sep}). The second set of data was much better since the target was in the centre of the PC and the exposure time was 1700s instead of 300s, spread over several exposures and allowing to reject cosmic rays easily. The accuracy of the result is then very good. The differences of magnitude in both filters indicate that the two components are probably very similar : $\Delta m_{F814W}= 0.36\pm0.07$ mag and $\Delta m_{F675W}= 0.53\pm0.09$ mag.

DENISPJ122813.8-154711 has been observed now over more than 5 years, and we have astrometric data related to the binary spread over more than 3 years. Within a few years we should thus be able to compute orbital parameters and dynamical masses. The motion of DENISPJ122813.8-154711-B is already obvious. We see in Table \ref{table_d1228_d0205} a separation change by $\sim 8\%$ and a position angle change by $23$ degrees in three years. If we assume a semimajor axis of 6.6 A.U, and a total mass of $0.12$ M$_{\sun}$, we find an orbital period of $\sim$ 49 years.

\subsection{2MASSW1239272+551537}
\objectname[2MASSW1239272+551537]{2MASSW1239272+551537} is one of the new binary candidates we report in this paper. It has been observed on the 18th of March 2001 in the F814W filter. We measure a separation of $211 \pm 2.8$ mas and a position angle of $187.6 \pm 0.3$ degrees. The difference of flux between the primary and the secondary is $\Delta m_{F814W}=0.34\pm0.07$ mag and $\Delta m_{F1042M}=0.54\pm0.09$ mag, suggesting that the two components might have very similar masses. The whole system is classified as L5 by \citet{kirk00}, who also evaluated the photometric distance of the whole system considering it was a single object (17 pc). From our photometric measurement we calculate a distance (corrected for the multiplicity) of 21.3 pc. Assuming a semi-major axis of 5.9 A.U, with a total mass of $\sim0.20$ M$_{\sun}$ (masses are here unknown and we assume masses of about $\sim$ 0.1 M$_{\sun}$ for each component), the period of this system would be $\sim$ 32.0 years.

\subsection{2MASSW1311392+8032222}
\objectname[2MASSW1311392+8032222]{2MASSW1311392+8032222} is a M8.0 dwarf \citep{gizis00} and is also one of the new binaries we report in this paper. It has been observed on the 30th of July 2000 in the F1042M and F814W filters. We measure a separation of $300 \pm 2.8$ mas and a position angle of $167.3 \pm 0.3$ degrees. The difference of flux between the primary and the secondary is $\Delta m_{F1042M}=0.45\pm0.09$ mag and $\Delta m_{F814W}=0.0.39\pm0.07$ mag, suggesting that the two components have similar masses.

We evaluate a photometric distance corrected for binarity of 13.7 pc. Assuming a semi-major axis of 5.5 A.U, with a total mass of $\sim0.20$ M$_{\sun}$ (masses are here unknown and we assume masses of about $\sim$ 0.1 M$_{\sun}$ for each component), the period of this system would be $\sim$ 28.8 years.

\subsection{2MASSW1426316+155701}
\objectname[2MASSW1426316+155701]{2MASSW1426316+155701} is a M9 dwarf and has been reported as a binary for the first time by \citet{close02a, close02b} with Hokupa'a/Gemini. On the 22nd of September 2001, they measured a separation of $152 \pm 6$ mas and a position angle of $344.1 \pm 0.7$ degrees. They estimated an age of $0.8^{+6.7}_{-0.2}$ Gyr and a photometric distance of $23.6 \pm 6.0$ pc, giving a separation of $3.6 \pm 0.9$ A.U and a period of $17^{+10}_{-7}$ years. They also estimated the mass of the whole system $M_{tot}=0.140_{-0.026}^{+0.011}$ $M_{\sun}$ as well as the masses of each components : $M_{A}=0.074_{-0.005}^{+0.011}$ $M_{\sun}$ at the limit of the Hydrogen burning limit for the primary and $M_{B}=0.066_{-0.006}^{+0.015}$ $M_{\sun}$ for the secondary.

On the 17th of July 2001 we measured a separation of $157.1 \pm 2.8$ mas and a position angle of $339.9 \pm 0.3$ degrees. We also estimate a photometric distance corrected for binarity of 26.7 pc. These values are close to the one of \citet{close02b} taken $\sim$2 months later (within 1$\sigma$). \citet{gizis00} measured a proper motion of $\mu=$0\farcs121/yr, implying 20 mas of motion between the observations, confirming that the system is a common proper motion pair. We also measured differences of magnitudes of $\Delta m_{F814W}=1.40\pm0.09$ mag, and $\Delta m_{F1042M}=1.30\pm0.09$ mag. Assuming a total mass of 0.14$M_{\sun}$ as given by \citet{close02a, close02b} and a semi-major axis of 4.3 A.U, we estimate the period of this system to be 33.3 years.

\subsection{2MASSW1430436+291541}
\objectname[2MASSW1430436+291541]{2MASSW1430436+291541} is one of the 15 new binary candidates we report in this paper. It has been observed on the 19th of April 2001 in the F1042M and F814W filters. We measure a separation of $88 \pm 2.8$ mas and a position angle of $327.5 \pm 1.2$ degrees. The difference of flux between the primary and the secondary is $\Delta m_{F1042M}=0.76\pm0.11$ mag and $\Delta m_{F814W}=0.78\pm0.11$ mag, suggesting that the two components have slightly similar masses. The companion appears clearly after PSF fitting, as shown in Figure \ref{close_bin}.

We evaluated the distance (corrected for multiplicity) of 29.4 pc. Assuming a semi-major axis of 3.4 A.U, with a total mass of $\sim0.20$ M$_{\sun}$ (masses are here unknown and we assume masses of about $\sim$ 0.1 M$_{\sun}$ for each component), the period of this system would be $\sim$ 13.9 years.

\subsection{DENISPJ144137.3-094559}
\objectname[DENISPJ144137.3-094559]{DENISPJ144137.3-094559} is a binary brown dwarf of spectral type L1 \citep{martin99a}, first resolved in February 2000 with Keck/NIRC (Mart\'\i n, E., private communication). It has been observed with Keck by \citet{stephens01} on the 13th of April 2000 and resolved with a separation of 0.42 arcsec. \citet{kirk00} deduced a distance of 25.5 pc using photometric measurements, considering a single object. We estimate a photometric distance corrected for multiplicity of 29.2 pc.

For this object we had two epochs of data, one from our HST program (two filters) and one from an on going program (GO9157).

In the first set of data obtained on the 16th of January 2001 we measured a separation of $375 \pm 2.8 $ mas and a position angle of $290.3 \pm 0.3$ degrees. The values obtained in both filters are again in very good agreement (cf. Figure \ref{pa_vs_sep}). We measured $\Delta m_{F675W}=0.63\pm0.09$ mag and $\Delta m_{F814W}=0.37\pm0.07$ mag.
In the second set of data, obtained four months later on the 22nd of May 2001, we measured a separation of $370 \pm 2.8$ mas, a position angle of $290.8 \pm 0.3$ degrees, and a difference of magnitude $\Delta m_{F814}=0.30\pm0.07$. These data are in good agreement with the previous one.
Assuming a semi-major axis of 14.1 A.U, with a total mass of $\sim0.2$ M$_{\sun}$ (masses are here unknown and we assume masses of about $\sim$ 0.1 M$_{\sun}$ for each component), the period of this system would be $\sim$ 118 years.

\subsection{2MASSW1449378+235537}
\objectname[2MASSW1449378+235537]{2MASSW1449378+235537} is another of the new binary candidates we report in this paper. The spectral type of the whole system is L0 \citep{kirk00}. It has been observed on the 21st of December 2000 in the F814W and F1042M filters. We measure a separation of $134 \pm 2.8$ mas and a position angle of $64.4 \pm 1.2$ degrees. The difference of flux between the primary and the secondary is $\Delta m_{F814W}=1.51\pm0.11$ mag and  $\Delta m_{F1042M}=1.08\pm0.11$ mag, suggesting that the secondary is redder than the primary.

\citet{kirk00} evaluated the photometric distance of the unresolved  system to 62 pc. From our photometric measurements we estimate a photometric distance corrected for binarity of 63.7 pc. Assuming a semi-major axis of 11.0 A.U, with a total mass of $\sim0.20$ M$_{\sun}$, the period of this system would be $\sim 81.1$  years.

\subsection{2MASSW1600054+170832}
\objectname[2MASSW1600054+170832]{2MASSW1600054+170832} is one of the 15 new binary candidates we report in this paper. It is a L1.5 dwarf according to \citet{kirk00}. It has been observed on the 14th of January 2001 in the F814W and F1042M filters. We measure a separation of $57 \pm 2.8$ mas and a position angle of $156.2 \pm 1.2$ degrees. The difference of flux between the primary and the secondary is $\Delta m_{F814W}=0.69\pm0.11$. At such a close separation, it might be better to consider 2- or 3-$\sigma$ uncertainties on these values, since we reach here the very limit of resolution of the WFPC2/PC camera. Nevertheless, the companion appears once again clearly after PSF subtraction (see Figure \ref{close_bin}). The object was too faint for PSF fitting in the F1042M filter.

\citet{kirk00} evaluated the photometric distance of the unresolved system to 62 pc. We estimate a photometric distance (corrected for multiplicity) of the primary of 60.6 pc. Assuming a semi-major axis of 4.3 A.U, with a total mass of $\sim0.20$ M$_{\sun}$ (masses are here unknown and we assume masses of about $\sim$ 0.1 M$_{\sun}$ for each component), the period of this system would be $\sim$ 19.9 years.

\subsection{2MASSW1728114+394859}
\objectname[2MASSW1728114+394859]{2MASSW1728114+394859} is also one of the new binary candidates. The spectral type of the whole system is L7 \citep{kirk00}. It has been observed on the 12th of August 2000 in the F814W and F1042M filters. We measure a separation of $131.3 \pm 2.8$ mas and a position angle of $27.6 \pm 1.2$ degrees. The difference of flux between the primary and the secondary in this range of wavelength is $\Delta m_{F814W}=0.66\pm0.11$mag, suggesting that the two components have similar masses. This object was also too faint for PSF fitting in the F1042M filter.

\citet{kirk00} evaluated the photometric distance of the unresolved system (20 pc). We estimate the photometric distance (corrected for multiplicity) at 20.4 pc. Assuming a semi-major axis of 3.4 A.U, with a total mass of $\sim0.20$ M$_{\sun}$, the period of this system would be $\sim$ 13.8 years.

\subsection{2MASSW2101154+175658}
\objectname[2MASSW2101154+175658]{2MASSW2101154+175658} is a L7.5 dwarf according to \citet{kirk00} and is a new binary we report in this paper. It has been observed on the 7th of May 2001 in the F814W filter. We measure a separation of $234.3 \pm 2.8$ mas and a position angle of $107.7 \pm 0.3$ degrees. The difference of flux between the primary and the secondary is $\Delta m_{F814W}=0.59\pm0.09$ mag, suggesting that the two components have similar masses. The object was too faint in the F1042M filter to be able to use the PSF fitting.

\citet{kirk00} evaluated its distance at 29 pc, considering a single object. We estimate the corrected photometric distance to be 23.2 pc. Assuming a total mass of $\sim0.20$ M$_{\sun}$ (masses are here unknown and we assume masses of about $\sim$ 0.1 M$_{\sun}$ for each component), and a semi-major axis of 7.1 A.U, the period of this system would be $\sim$ 41.9 years.

\subsection{2MASSW2140293+162518}
\objectname[2MASSW2140293+162518]{2MASSW2140293+162518} has been reported as a binary for the first time by \citet{close02a, close02b} with Hokupa'a/Gemini. On the 22nd of September 2001, they measured a separation of $155 \pm 5$ mas and a position angle of $134.3 \pm 0.5$ degrees. They estimated an age of $3.0^{+4.5}_{-2.4}$ Gyr and a photometric distance of $23.9 \pm 6.0$ pc, leading to a separation of $3.7 \pm 0.9$ A.U and a period of $18^{+10}_{-7}$ years. They also estimated the mass of the whole system $M_{tot}=0.162_{-0.018}^{+0.008}$ $M_{\sun}$ as well as the masses of each components : $M_{A}=0.087_{-0.017}^{+0.008}$ $M_{\sun}$ for the primary and $M_{B}=0.075_{-0.018}^{+0.007}$ $M_{\sun}$ for the secondary, at the limit of the Hydrogen burning limit. From their colours they estimated the spectral types of each component : M$8.5 \pm 1.5$ for the primary and L$0 \pm 1.5$ for the secondary.

On this object we had data from the HST public archive (GO8581, P.I Reid). The object was observed on the 21st of May 2001, in the PC chip of WFPC2 in two filters (F814W and F1042M), almost four months before \citet{close02a} reported it. On this set of data we measure a separation of $159.0 \pm 2.8$ mas and a position angle of $132.4 \pm 0.3$ degrees, as well as difference of magnitude of $\Delta m_{F814W}=1.51\pm0.11$ mag, and $\Delta m_{F1042M}=1.38\pm0.11$ mag. These values are in agreement with the values given by \citet{close02b} within 3$\sigma$. We estimate a photometric distance of 12.9 pc. Considering a semi-major axis of 2.6 A.U and a total mass of 0.16 $M_{\sun}$ as calculated by \citet{close02b}, the period of this system would be 10.5 years.
\citet{gizis00} measured a proper motion of $\mu_{\alpha}=$-0\farcs008/yr and $\mu_{\delta}=$-0\farcs102/yr implying 35 mas of motion between the observations. Although the uncertainties on their astrometry are relatively high, the system is likely to be a common proper motion pair. Better astrometric measurements are required in order to confirm the multiplicity of this object.

\subsection{2MASSW2147437+143131}
\objectname[2MASSW2147437+143131]{2MASSW2147437+143131} is a M8.0 ultra-cool dwarf \citep{gizis00} and also one of the new binaries we report in this paper. It has been observed on the 9th of October 2000 in the F1042M and F814W filters. We measure a separation of $322 \pm 2.8$ mas, a position angle of $329.5 \pm 0.3$ degrees and a difference of magnitude in the F814W filter of $\Delta m_{F814W}=0.62\pm0.09$ mag. We were not able to measure accurately the parameters in the F1042M filter due to a a cosmic ray event very close to the primary. We were only able to evaluate a difference of magnitude of $\Delta m_{F1042M}=0.75\pm0.28$ mag.

The photometric distance given by \citet{gizis00} is 22.9 pc considering a single object. From our measurements we estimate a photometric distance corrected for multiplicity of 21.8 pc. Thus assuming a total mass of $\sim0.20$ M$_{\sun}$ (masses are here unknown and we assume masses of about $\sim$ 0.1 M$_{\sun}$ for each component), and a semi-major axis of 9.1 A.U, the period of this system would be $\sim$ 61.7 years.

\subsection{2MASSW2206228-204705}
\objectname[2MASSW2206228-204705]{2MASSW2206228-204705} has also been reported as a binary for the first time by \citet{close02a, close02b} with Hokupa'a/Gemini. On the 22nd of September 2001, they measured a separation of $168 \pm 7$ mas and a position angle of $68.2 \pm 0.5$ degrees. They estimated an age of $3.0^{+4.5}_{-2.4}$ Gyr and a photometric distance of $24.68 \pm 6.8$ pc, giving a separation of $4.1 \pm 1.1$ A.U and a period of $20^{+9}_{-7}$ years. They also estimated the mass of the whole system $M_{tot}=0.178_{-0.014}^{+0.008}$ $M_{\sun}$ as well as the masses of each components : $M_{A}=0.090_{-0.014}^{+0.008}$ $M_{\sun}$ for the primary and $M_{B}=0.088_{-0.014}^{+0.008}$ $M_{\sun}$ for the secondary. From their colours they estimated the spectral types of each component : M$8.0 \pm 1.5$ for the primary and M$8.5 \pm 1.5$ for the secondary. \citet{kirk00} assigned M8 to the whole system.

We used again data from the HST public archive (GO8581, P.I Reid). The object was observed on the 13th of August 2000, in the PC chip of WFPC2 in two filters (F814W and F1042M), almost fourteen months before \citet{close02a} reported it. On this set of data we measure a separation of $160.7 \pm 2.8$ mas and a position angle of $57.2 \pm 0.3$ degrees, as well as difference of magnitude of $\Delta m_{F814W}=0.36\pm0.07$ mag, and $\Delta m_{F1042M}=0.30\pm0.07$ mag. 

From our photometric measurement, we estimate a distance (corrected for multiplicity) of 22.2 pc. Considering a semi-major axis of 4.5 A.U and a total mass of 0.178 $M_{\sun}$ the period of this system would be 22.6 years. If we consider the little motion we observe between the two epochs (11 degrees in 14 months), and assuming a circular face-on orbit, it gives a period of 33 years which is of the same order of the previous estimation (and in good agreement since the orbit is probably not circular face-on). 

\citet{gizis00} measured a proper motion of $\mu=$0\farcs065/yr, implying 76 mas of motion between the observations. Although the uncertainties on their astrometry are relatively high, the system is likely to be a common proper motion pair. Better astrometric measurements are required in order to confirm the multiplicity of this object.

\subsection{2MASSW2331016-040618}
\objectname[2MASSW2331016-040618]{2MASSW2331016-040618} has also been reported as a binary for the first time by \citet{close02a, close02b} with Hokupa'a/Gemini. On the 22nd of September 2001, they measured a separation of $573 \pm 8$ mas and a position angle of $302.6 \pm 0.4$ degrees. They estimated an age of $5.0^{+2.5}_{-4.4}$ Gyr and a photometric distance of $25.2 \pm 6.8$ pc, thus giving a separation of $14.4 \pm 3.9$ A.U and a period of $139^{+86}_{-57}$ years. They also estimated the mass of the whole system $M_{tot}=0.153_{-0.020}^{+0.008}$ $M_{\sun}$ as well as the masses of each components : $M_{A}=0.091_{-0.013}^{+0.008}$ $M_{\sun}$ for the primary and $M_{B}=0.062_{-0.020}^{+0.010}$ $M_{\sun}$ for the secondary, thus clearly below the sub-stellar limit. From their colours they estimated the spectral types of each component : M$8.0 \pm 1.5$ for the primary and L$3 \pm 1.5$ for the secondary. This object is very peculiar because the differences of magnitudes between the primary and the secondary are really high in the infrared. \citet{close02b} measured $\Delta K=2.38 \pm 0.16$ mag.

We used data from the HST public archive (GO8581, P.I Reid). The object was observed on the 6th of May 2001, in the PC chip of WFPC2 in two filters (F814W and F1042M), almost 4.5 months before \citet{close02a} reported it. On this set of data we measure a separation of $577 \pm 2.8$ mas and a position angle of $293.7 \pm 0.3$ degrees, as well as difference of magnitude of $\Delta m_{F814W}=3.90\pm0.09$ mag, and $\Delta m_{F1042M}=3.54\pm0.09$ mag.  The results we obtain in the two filters are slightly different but still in agreement within 3$\sigma$ (cf. Table \ref{results}). Our value for the separation is in good agreement with the values given by \citet{close02b} but the difference in the values for the position angle is larger. Even if the difference of magnitude is high, the separation is large enough to let us trust the results we obtained with our PSF fitting method with the accuracy of 0\farcs0028 for the separation and 0.3$^{\circ}$ for the P.A, as explained in section \ref{dataanalysis}.

\citet{gizis00} measured a proper motion of $\mu_{\alpha}=$0\farcs401/yr and $\mu_{\delta}=$-0\farcs231/yr, implying a motion of 0\farcs150 in right ascension and -0\farcs087 in declination between the observations, whereas the variation of position angle occurs in the opposite direction, confirming that the system is a common proper motion pair. 

On the basis of distance and proper motions consistencies, \citet{gizis00} suggested that 2MASSW2331016-040618 is associated to the F8 dwarf \objectname[HD221356]{HD221356}. The distance (26.2 pc) and proper motion ($\mu_{\alpha}=178.65$ mas/yr and $\mu_{\delta}=-192.79$ mas/yr) of HD221356 have been measured precisely by Hipparcos. The distances of the two objects are very similar (26.2 pc for the M-dwarf and 26.24 for the F-dwarf) but the proper motions measured by \citet{gizis00} and Hipparcos \citep{perryman97} are very different : more than 220 mas/yr of difference for $\mu_{\alpha}$ and almost 40 mas/yr for $\mu_{\delta}$. We then suggest that these two objects are not linked on the basis of proper motion discrepancy. HD221356 does not appear on the HST images we used, thus not allowing to check for second epoch data. It might nevertheless be interesting to get second epoch data in the future to confirm the values of proper motion of 2MASSW2331016-040618 given by \citet{gizis00}. We therefore consider that 2MASSW2331016-040618 is a field binary ultra-cool dwarf.

We estimate a photometric distance corrected for the binarity of 26.2 pc. Considering a semi-major axis of 19.0 A.U, and a total mass of 0.153 $M_{\sun}$, the period of this system would be 211 years.

\subsection{SDSS2335583-001304}
\objectname[SDSS2335583-001304]{SDSS2335583-001304} is a new very close binary system candidate. We measured a separation of $56.8 \pm 2.8 $ mas and a position angle of $\sim 8.1 \pm 1.2$ degrees. The system is so close that it was not resolved in the F1042M filter. We are here very close to the HST/WFPC2 pixel scale and already below the sampling limit, thus second epoch data with higher resolution will be necessary in order to confirm the multiplicity of this object. The difference of magnitude in the F814W is : $\Delta m_{F814W}=1.00\pm0.11$ mag. This value should be considered with caution, since we reach here the limits of resolution of HST/WFPC2. It might thus be better to consider 2- or 3-$\sigma$ uncertainties.

Nevertheless the clear elongation of the object in the image as well as the companion that appears clearly after PSF fitting as shown in Figure \ref{close_bin} allow us to conclude that SDSS2335583-001304 is a good candidate multiple system. 

Unfortunately the only photometric value available is the one we measured, thus not allowing to compute the photometric distance.

Assuming a circular face-on orbit at a distance of 20 pc, with a total mass of $\sim0.20$ M$_{\sun}$ (masses are here unknown and we assume masses of about $\sim$ 0.1 M$_{\sun}$ for each component), and a radius of 1.4 A.U, the period of this system would be $\sim$ 3.7 years only. At a distance of 40 pc, this period would be 10.9 years. This makes of SDSS2335583-001304 one of the best candidates for a direct estimate of dynamical masses within a few years only. 

\section{Analysis}
\subsection{The Binary Fraction \label{bin_freq}}
26 of the 134 targets in the sample appear to be multiple systems (see Figure \ref{all_bin}). Fifteen of these still have to be confirmed by second epoch data, but given the relatively low star density, we will consider that these objects are almost certainly binary systems. 2MASSW1112256+354813 is associated to a G-dwarf in a triple system, and thus is not a field binary and we therefore remove it from the statistics. We then hereafter count 25 field binaries. Eleven binary systems were already known and thus confirmed by second and sometimes third epoch data. 

\begin{figure}
\plotone{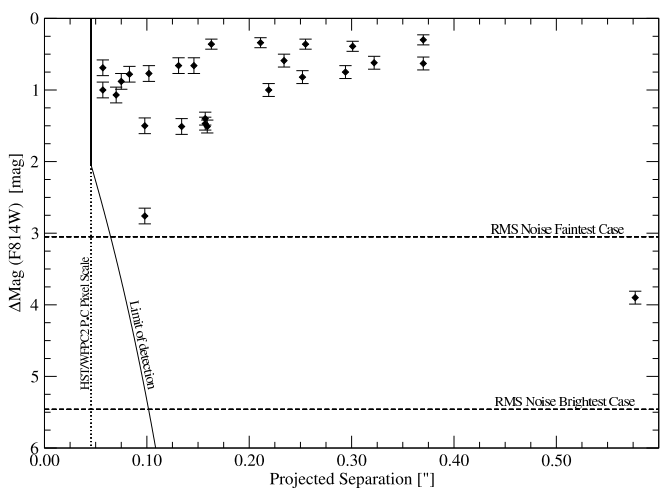}
\caption{Limits of detection. The largest difference of magnitude we were able to detect is shown here as a function of the angular separation. At separations close to the pixel scale, the detection of companions is limited by the contrast against the bright primary. At larger separations, the detection is background limited. The full line represents the HST/WFPC2 limit of detection and the 2 dashed lines the 3$\sigma$ RMS noise in the faintest and brightest cases.  The limit of detection has been computed by calculating the average of the 3$\sigma$ RMS noise on the radial profile of several unresolved objects. The results obtained for the 26 binaries reported in this paper are represented by filled diamonds with their uncertainties. \label{calib1}}
\end{figure}

\begin{figure}
\epsscale{0.8}
\plotone{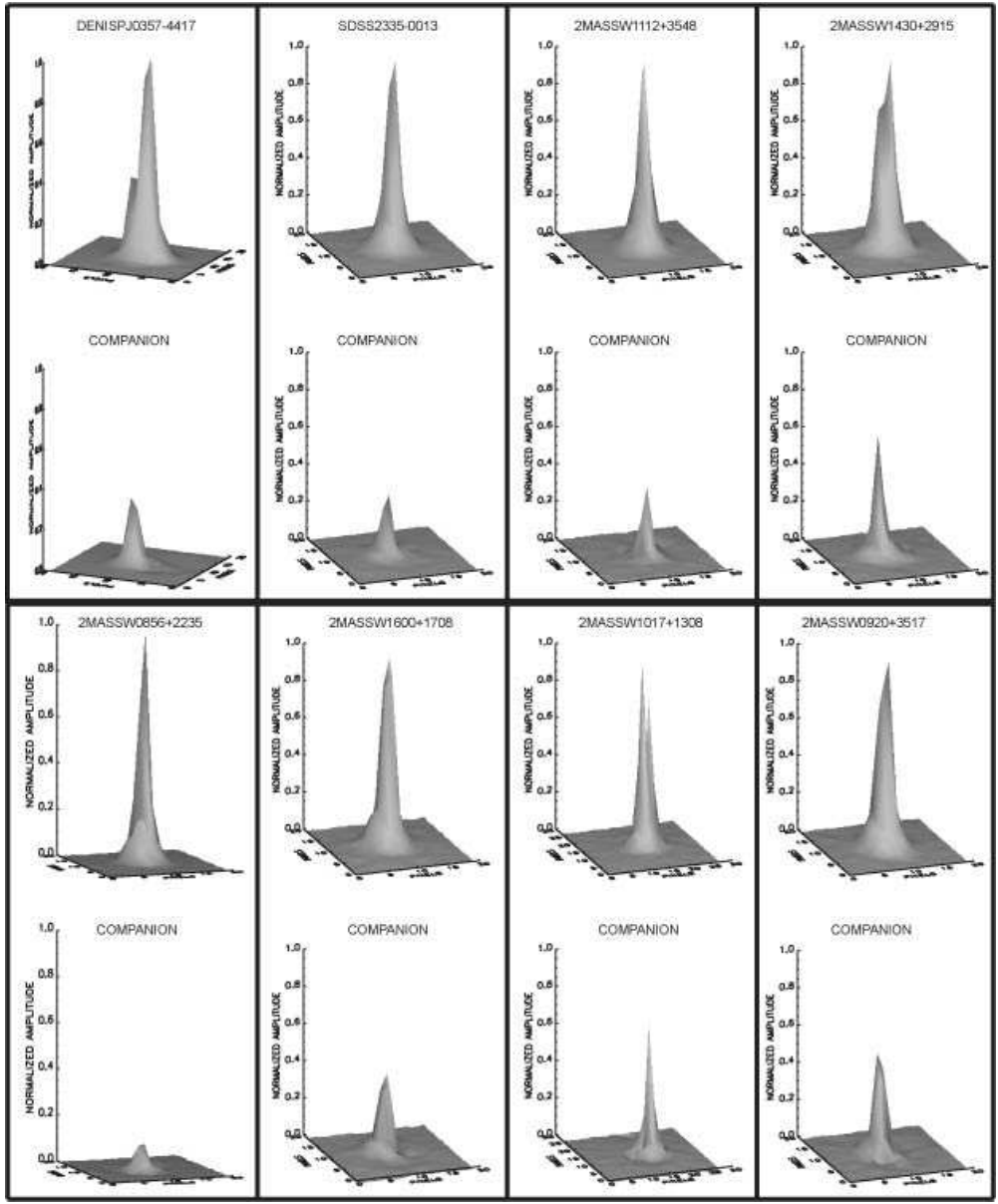}
\caption{Surface plots representing the closest binary candidates of the sample and their companion. The companion appears after subtraction of the primary using the PSF fitting. All these plots correspond to the F814W images. Amplitudes are normalized and sky has been subtracted. Even if the separations are very small, the companions appear clearly after the PSF subtraction. \label{close_bin}}.
\end{figure}

\begin{figure}
\epsscale{0.8}
\plotone{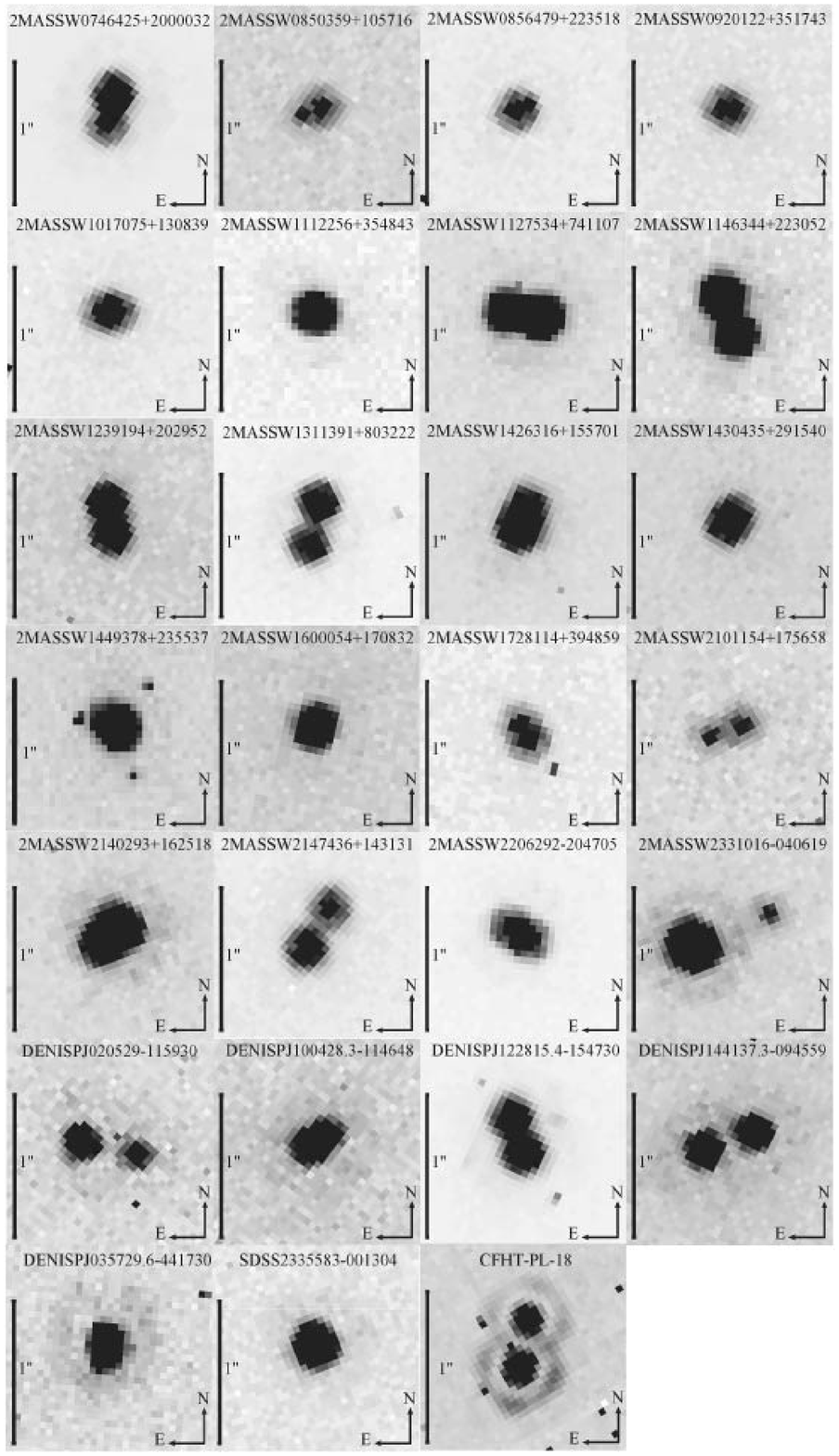}
\caption{Mosaic of the 26 binaries and binary candidates presented in this paper (HST/WFPC2, F814W filter) and CFHT-PL-18 (\citet{martin98}, HST/NICMOS, F165W filter)\label{all_bin}}.
\end{figure}

\begin{figure}
\plotone{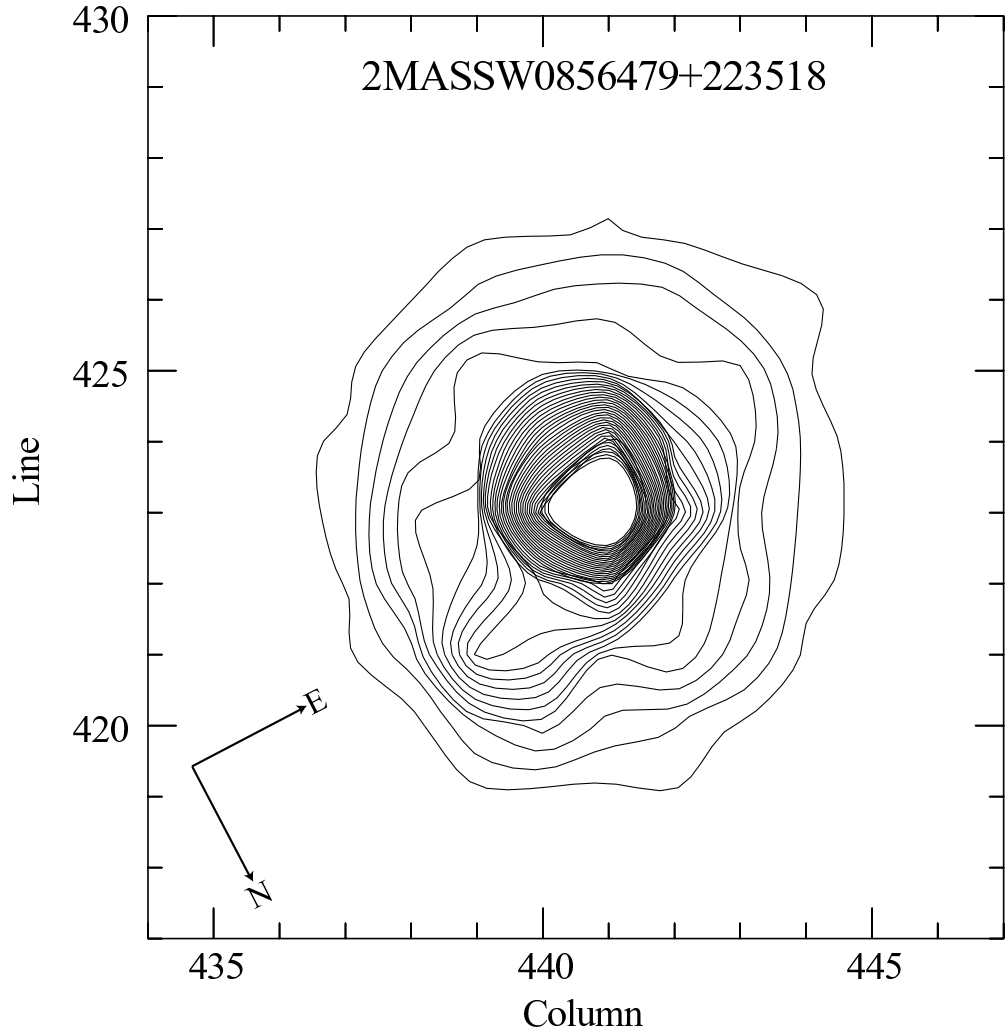}
\caption{Contour plot of 2MASS0856+2235 obtained with HST/WFPC2. We measured a position angle of $275 \pm 2.0$ degrees and a separation of 98$\pm$9 mas. The two centroids appear clearly. \label{contour_plot}}
\end{figure}

Our observed binary frequency is based on the fact that we know 25 binaries in a sample of 133 ultra-cool dwarfs. This corresponds to an observed binary fraction of $18.8 \pm 3.7 \%$. This value is only indicative and not related to any physical value for the following reasons:
\begin{itemize}
\item  First because our detection was limited to binaries with differences of magnitude $\Delta m_{F814W}$ less than 3 mag (5.5 mag in the best cases, $\sim$4 mag in average, cf. Figure \ref{calib1}). We thus missed all the multiple systems for which the companion was too faint in comparison with the primary. Nevertheless, as shown in Figures \ref{calib1} and \ref{distri_dmag} and as explained in section \ref{lum_ratio}, our observations suggest a strong preference for nearly equal luminosity systems. We therefore go on the assumption that multiple systems with high differences of magnitudes are rare among ultra-cool dwarfs and that we do not miss a significant number of such multiple systems in our study except if there is a bimodal distribution of mass ratio with a second peak at small mass ratio.

\item Second because the sample selection introduced a bias, since we did
  not choose the targets completely randomly among field brown dwarfs: some
  were selected from their colours and brightnesses. The sample is thus
  limited in magnitudes rather than in distances, and this introduces a
  bias: the multiple systems are detected at larger distances than simple
  one, and their spatial density is thus overestimated. To be able to
  evaluate more precisely the binary fraction we have been analysing the
  sample in order to find at which limiting distance the sample cannot be considered as a complete volume-limited sample. As the sample is made of objects
  coming from different surveys (DENIS, 2MASS ans SDSS), we had to take
  their respective limits of completeness in account. The DENIS survey is
  complete until objects with I$\leq$18 mag \citep{delfosse99}. For a typical ultra-cool dwarf of spectral type L5 (latest DENIS spectral type of the sample, except DENIS0205-1159), it
  corresponds to an absolute magnitude of M$_{I}\sim$ 17 mag (according to
  the M$_{I}$ vs. Spectral Type relation given in \citet{dahn2002}), and to
  a distance of $\sim$ 16 pc. We consider for our study that the 2MASS
  survey has a limit of completeness of K$\leq$14.5 mag. For a typical
  ultra-cool dwarf of spectral type L8, this corresponds to an absolute
  magnitude M$_{K} \sim$13 mag, and to a distance of $\sim$20 pc. We can
  therefore as a first approximation consider that the sample was very
  close to a volume limited sample until 16 pc for the DENIS objects and
  until 20 pc for the 2MASS objects. As the SDSS objects represent only a
  very small number of objects in the sample, as many of them were also
  observed in the 2MASS survey and as they are all further away than 20 pc,
  they do not count in the statistical study hereafter. Figure
  \ref{distances} shows that on the range 0$\sim$20 pc (distance corrected
  for the resolved multiple systems), the sample behaves almost like a
  volume limited sample and the number of objects increases like
  $d^{3}$. 

  After 20 pc, the sample behaves
  indeed more like a magnitude limited sample and the number of objects
  decreases roughly like $d^{-2}$. All these considerations suggest to us that
  in order to make a better statistical study, we should consider only the
  DENIS objects of the sample that are closer than 16 pc and the 2MASS
  objects that are closer than 20 pc. Only one DENIS object (no binary) and
  twenty five  2MASS objects (including four multiple systems) are thus included
  in the statistical sample defined as explained above. This makes a binary
  fraction of $\sim 15\%$. 

\item Finally because the detections are limited to objects with separations greater than $\sim$ 0.060 arcsec. We therefore missed all the binaries with separations less than this value. This effect becomes stronger with increasing distances. At a distance of 20 pc, it corresponds to physical separations of 1.2 A.U. The value given above is then probably a lower limit of the exact binary fraction. 

\end{itemize}

Another way to estimate the bias corrected binary frequency is to apply a
correction to the observed binary frequency as described in
\citet{burgasser03} (see their equations 4 and 5). If we go on the
assumption that the total sample is purely magnitude limited, and that the
correction factor for the increase of volume sampled for binaries is
between $1.86\leq \alpha \leq 2.83$ (see section \ref{lum_ratio}), the
corrected binary fraction is $7.6^{+5.9}_{-1.7}\%$. As we observe a
preference for nearly equal luminosity systems (see section
\ref{lum_ratio}), $\alpha$ must be closer to 2.83 than to 1.86. The value
given above therefore corresponds to $\alpha=$2.83, but its uncertainties
includes the uncertainties on $\alpha$ and on the observed binary
fraction. This result is of the same order than the previous one and in
good agreement within the uncertainties. 

However, the two methods presented here to correct the observed binary frequency take in account only the bias for visual binaries resolved by HST, and not for the spectroscopic binaries of the sample that we miss. Indeed in order to construct our ``volume limited'' sample we correct the photometric distances of the identified HST visual binaries, but not those of undetected spectroscopic binaries. The same is true for the \citet{burgasser03} method; in their equation 4,  only the volume sampled for HST visual binaries is corrected. As result, the denominator of the binary fraction, which is the total number of systems (single and multiple), is overestimated because the bias for multiplicity is not corrected for all the multiple systems, and the binary fraction is therefore under-estimated.

Nevertheless, this sample of 134 objects is the largest studied to date, with the highest spatial resolution available. We found 15 new binary candidates (to be added to the 11 previously known). This allows us to make a meaningful statistical study over a sufficiently large number of objects. Although it is biased, we can already make some preliminary comments on the results, keeping in mind the limitations of this study. 

With these considerations we obtain a HST visual binary fraction among ultra-cool dwarfs of $7-10\%$. But even in these conditions and with these precautions the sample is not a purely volume limited sample and the results will not be as accurate as the one we could get with a statistically well defined sample.

\subsection{Distribution of Separations}
Figure \ref{distri_sep} shows the distribution of separations. It shows evidence of a lack of binaries with separations wider than $\sim 15$ A.U. This was already mentioned as the possible ``brown dwarf wide binary desert'' by \citet{martin2000c}. 

This is certainly not an artifact of incompleteness in the observations since this separation corresponds to $\sim$ 0\farcs5 (at an average distance of $\sim 20$ pc), where multiple systems can easily be detected either by HST/WFPC2 for the faintest companions or 2MASS and DENIS in the brighter and wider cases (the resolutions of the 2MASS and DENIS surveys are $\sim$ 2\farcs0, corresponding to $\sim 40$ A.U semi-major axis at 20 pc). Figure \ref{distri_sep} shows clearly that all the objects have angular separations well below these limits. We estimate that we should have been able to detect every companion candidate within 4\farcs0 of the targets (within the limits of flux ratio mentioned in Section \ref{lum_ratio}). We have been looking for wide companions by comparing the colours of the objects in the WFPC2 field of view, and found only very few candidates with colours very similar to those of the associated targets. The colours similarity being a poor constraint (especially with the HST filters used in this study), the probability that these objects are physically associated to the corresponding targets is very low. One object from the literature (CFHT-Pl-18, cf. Table \ref{table_periods}) is currently known to have a wider separation (35 A.U), but even in this case the separation does not exceed a few tens of A.U. This indicates that these objects can exist but are probably very rare, and probably do not have separations greater than a few tens of A.U.

\subsection{Luminosity ratios \label{lum_ratio}}
\begin{figure}
\plotone{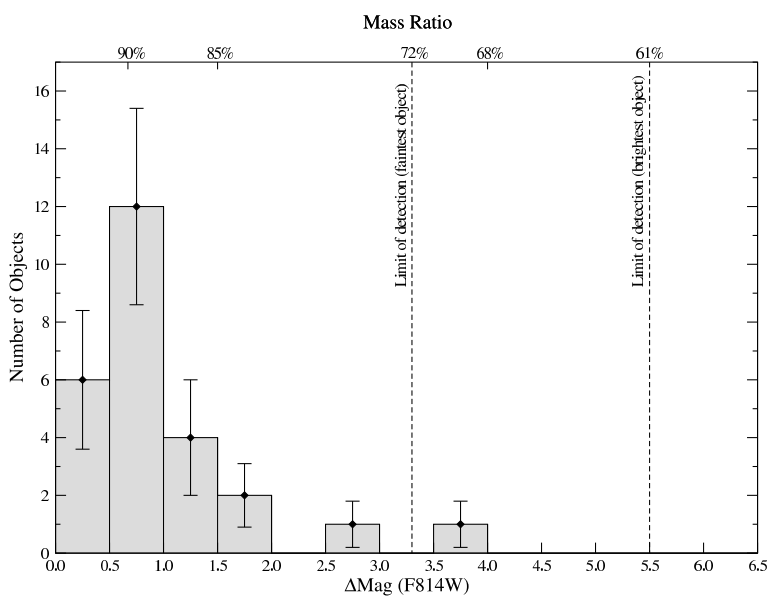}
\caption{Distribution of difference of magnitude in the F814W ($\sim$ I) Filter. Although we would have been able to detect objects with $\Delta m_{F814W}<3.3$ easily, we observe a clear lack of systems with $\Delta m_{F814W}>2.0$, corresponding to a mass ratio of $85\%$ (the mass ratios indicated here correspond to the one given by the \citet{chabrier00} models for a primary of 0.1 $M_{\sun}$ and an age of 1 Gyr, cf. Table \ref{model}). \label{distri_dmag}}
\end{figure}

\begin{figure}
\plotone{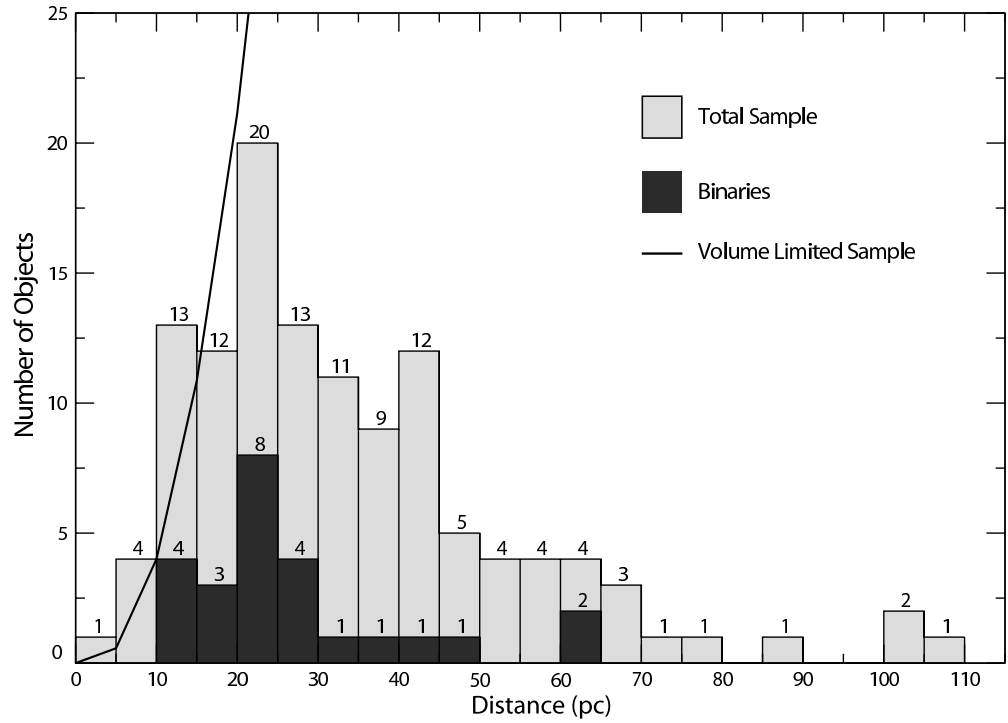}
\caption{Distribution of separation in the sample. The total sample is represented in light grey  and the binaries alone are over-plotted in dark grey. We also over-plot the expected distribution for a volume limited sample. The sample is roughly limited in magnitude. For the statistical study, we therefore consider only the DENIS objects at distances less than 16 pc and the 2MASS objects closer than 20pc. \label{distances}}
\end{figure}

\begin{figure}
\plotone{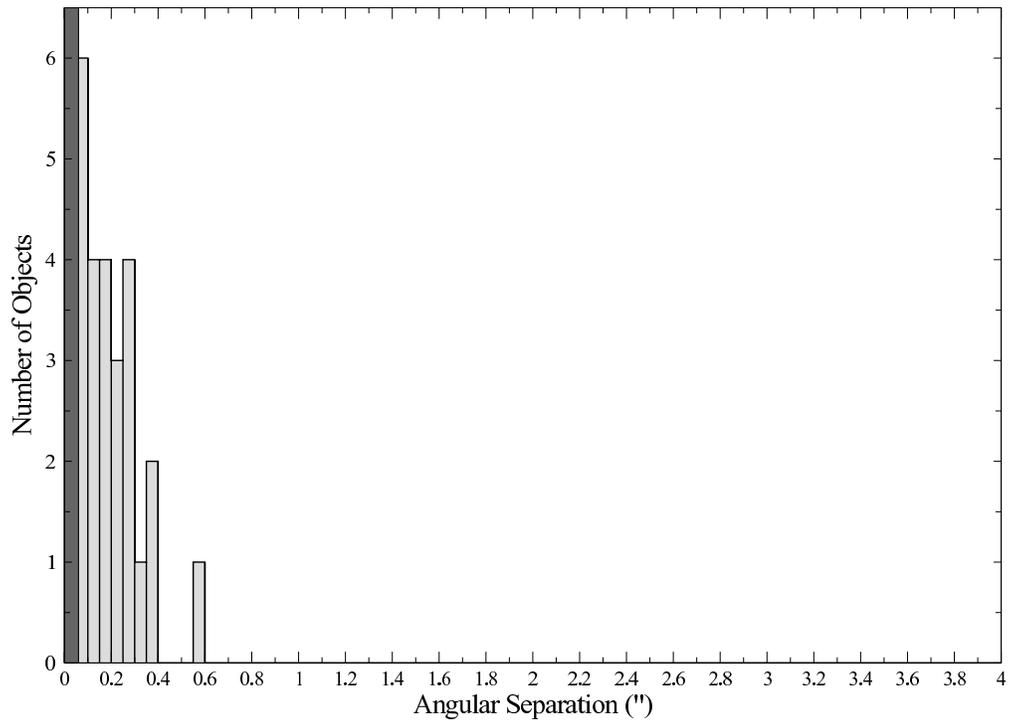}
\caption{Distribution of angular separations of the binaries presented in this paper. There is an evident absence of systems with separation larger than 0\farcs6. The observations were limited to 0\farcs060 in the lower part (dark grey region) and $\sim$ 4\farcs0 for the upper part. \label{distri_sep}}
\end{figure}

Figure \ref{calib1} and \ref{distri_dmag} also suggest that there is a lack of binary systems with large differences of magnitudes, corresponding to unequal luminosities. Most of the values of differences of magnitudes are homogeneously spread between $0.1 \lesssim \Delta m_{F814W} \lesssim 2.0$ mag. Only two systems have $\Delta m_{F814W} > 2.5$ mag. As explained in Section \ref{dataanalysis}, this is probably not an artifact of incompleteness, because even if the sample is magnitude limited (in a magnitude limited sample, the selection bias is stronger for the equal luminosity systems, since the systems with unequal luminosities are detected in a smaller volume) Figure \ref{distri_dmag} shows that we should have been able to detect all the binary systems with differences of magnitudes between $1.5 \lesssim \Delta m_{F814W} \lesssim 3.0$ mag easily even in the faintest cases, and binary systems with differences of magnitudes between  $3.0 \lesssim \Delta m_{F814W} \lesssim 5.5$ mag in the brightest cases. 

Binaries with a flux ratio $f=f_2/f_1$ can be detected $\sqrt{1+f}$ times further away. Therefore the higher ($\sim$ 1) the flux ratio is, the larger the sampled volume is. This is an another well known effect of the bias introduced by the multiplicity: the number of equal mass binaries is over-estimated in comparison with the number of binaries with large differences of magnitudes. This effect is maximum for equal mass binaries compared to binaries with mass ratio close to zero. The volume sampled for equal mass binaries is then 2.83 times larger than the volume sampled for systems with mass ratio close to zero. In our study we find a factor of $24/2=12$ times more systems with $0.1 \lesssim \Delta m_{F814W} \lesssim 2.0$ mag than system with $2.0 \lesssim \Delta m_{F814W} \lesssim 5.5$ (see Figure \ref{distri_dmag}). Such a very large ratio can not be explained by this bias.

\subsection{Colours \label{colours}}
\begin{figure}
\plotone{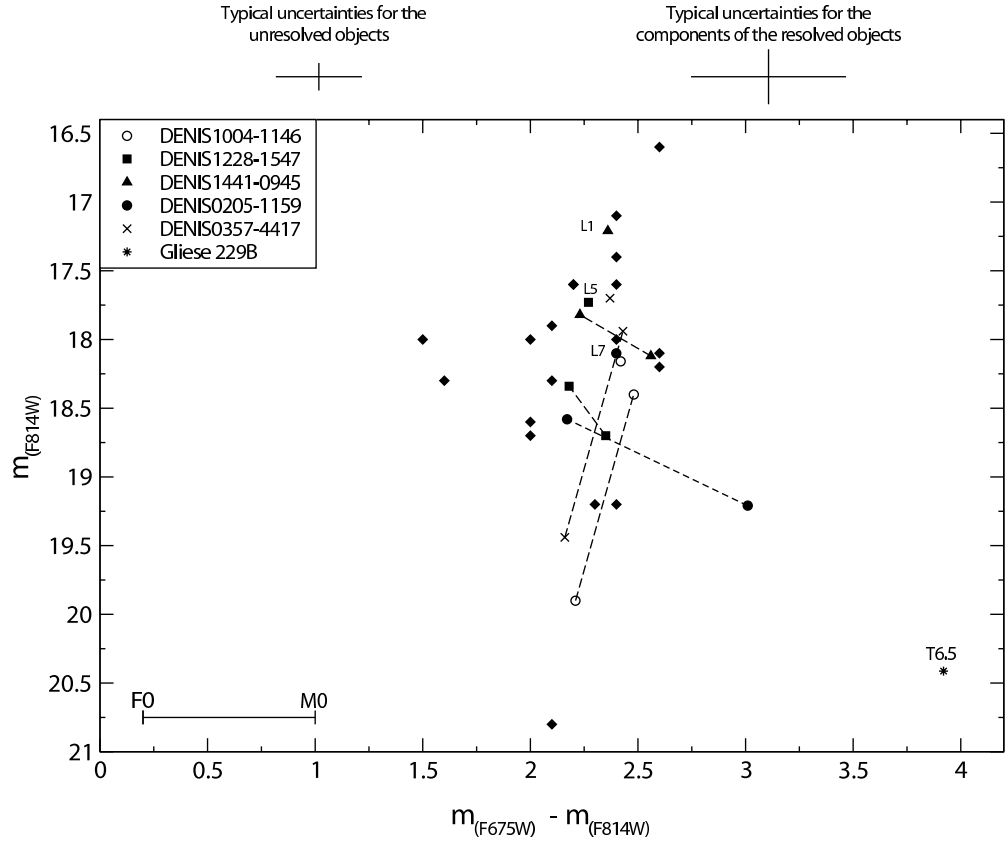}
\caption{\label{iri} Colour Magnitude diagram of the sample (GO8720). Single or unresolved objects are represented by filled diamonds. Binary systems are represented by various symbols. For each binary system, both the binary system and the two components are represented (same symbol), these  latter two are joined by a dashed line. The T-dwarf Gl229B is represented (by an asterisk) in addition for comparison between M, L and T-dwarfs. Values for Gl229B from \citet{golimowski98}. Not all objects were available in both filters (see Tables \ref{results} and \ref{photom_go8720}). Spectral types of some objects were available (cf. Table \ref{targets}).}
\end{figure}

A colour-magnitude diagram (F814W vs. (F675W-F814W) $\sim$ I vs. R-I) of the sample of the GO8720 program is given in Figure \ref{iri}. Plotted are the single or unresolved targets, as well as the multiple systems represented by three points : one point for the whole system and two linked points for the two individual components. Figure \ref{iri} shows clearly that the multiple systems include some of the reddest objects of the sample. The slopes of the lines joining the primary and the secondary might indicate that for two objects the secondary is bluer (higher R-I), whereas for the three other it is redder. This might be explained by dust effects as described in the DUSTY model \citep{allard01, chabrier00}, but could also be due to the relatively high uncertainties on the flux ratios and the corresponding uncertainties on the magnitudes of the two components.

\begin{figure}
\plotone{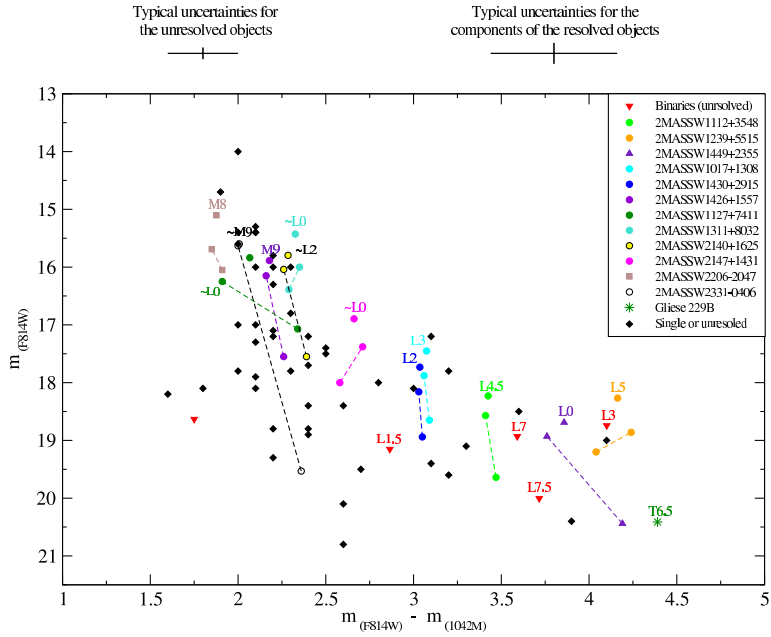}
\caption{\label{izi} Colour Magnitude diagram of the sample (GO8581). Some multiple systems were unresolved or too faint in the F0142M filter, thus not allowing to measure the flux ratios. These latter cases are represented by a red triangle. The binaries resolved in both filters are represented with 3 points like in Fig. \ref{iri}. Typical uncertainties are given for the unresolved objects and for the components of the multiple systems. The T-dwarf Gl229B is represented (asterisk) in addition for comparison between late M, L and T-dwarfs. Values for Gl229B from \citet{golimowski98}. (see Tables \ref{results} and \ref{photom_go8581} for the detailed values of magnitudes)}
\end{figure}

A colour-magnitude diagram (F814W vs. (F814W-F1042M) $\sim$ I vs. I-z) of the sample of the GO8581 program is given in Figure \ref{izi}. Not all the 15 binaries found in this sample were bright enough in the F1042M filter to be able to do the PSF fitting, thus not allowing to measure the flux ratios. 

\section{Discussion}

This study allows us to point out three important results: the frequency of binaries, the lack of binaries with separations greater than 15 A.U and the possible preference for equal-mass systems. We will now discuss these results in the context of current scenarios of formation and evolution of free-floating ultra-cool dwarfs.

\subsection{Binary frequency}
\begin{figure}
\plotone{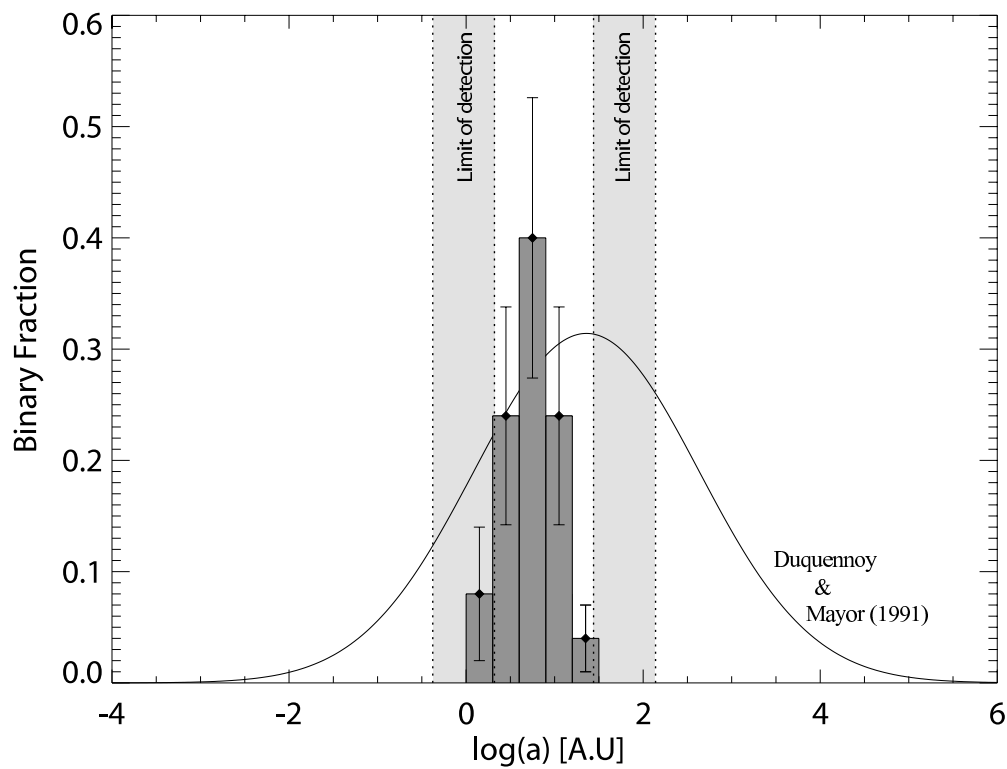}
\caption{Distribution of separations of the 25 binaries presented in this paper, compared with the \citet{duquennoy91} distribution of G-dwarf. The upper and lower limits of detection are in the grey regions. These regions have been calculated considering an upper limit in the distance at 35 pc (average distance of the sample) with corresponding maximum limits of detection of 4\farcs0 and 0\farcs06 and a lower limit at 7 pc (closest distance of our sample) with minimum limits of detection of 4\farcs0 and 0\farcs060. Even if it is not correct to compare both distributions (the distribution of ultra-cool binaries was not corrected for bias) the lack of binaries wider than 15 A.U and the strong peak around 2$\sim$4 pc appear clearly, although wider binaries should have been identified despite the selection biases. \label{histo}}
\end{figure}

The value we found for the binary frequency (10$\sim$15\%) is much lower than the values given by \citet{duquennoy91} for G dwarfs ($\sim 57\%$), and by \citet{reid97} or \citet{lydie02} for the early M dwarfs ($31-35 \%$) considering that this latter values covers larger ranges of mass ratios and separations whereas we were limited in both cases. If we compare with only the G and M stellar systems having a separation $2.1 \leq a \leq 140$ A.U (corresponding to respectively 0\farcs060 and 4\farcs0 at the average distance of our sample: 35 pc) we find roughly $\sim 30\%$ in the \citet{duquennoy91} distribution, which is still much higher than the one we give here for ultracool dwarfs (spectral types between M7 and L8). For comparison, in Figure \ref{histo} we combine a plot of the 25 field binary ultracool dwarfs presented here, with the distribution of separations given  by \citet{duquennoy91} for the G-dwarfs. In the left part of our histogram we missed the binaries with separations smaller than $\sim$ 0\farcs060. Even if it is not correct to compare directly both distributions (the \citet{duquennoy91} distribution was indeed corrected for biases whereas the distribution of the 25 binaries is not corrected at all), we can note the absence of systems with separations greater than $15$ A.U, and the strong peak between $4 \sim 8$ A.U, instead of $\sim 30$ A.U for the G dwarfs.  

In a recent study \citet{burgasser03} report an observed binary frequency among T dwarfs of $20^{+17}_{-7}\%$ (2 multiple systems over a sample of 20 objects) corresponding to a biased corrected fraction of $9^{+15}_{-4}\%$. This number is in good agreement with the one we report here for late-M and L dwarfs. They also computed a bias corrected value of the binary fraction of late-M and L dwarfs of the sample of \citet{reid01} (HST program GO8146, 4 multiple objects among 20 targets) and, assuming it was a purely magnitude limited sample, they estimated a fraction of binary of the same order ($9^{+11}_{-4}\%$) (see section \ref{bin_freq} for a detailed discussion on the difficulty to correct properly the bias). The results are again in good  agreement within $1\sigma$. Our binary fraction is based on a total sample 6.65 larger than the samples of T-dwarfs and L-dwarfs they present, including the sample of L-dwarfs of \citet{reid01}. 

The  binary frequency we report here gives also strong constraints on the models of formation. In particular such a high rate of multiple systems cannot be explained by the models of ejection \citep{reipurth01} that predict very few binary ultra-cool dwarfs. In a recent study \citet{delgado03, bate02} report that their models of formation and evolution predict a very low frequency of binaries among very low mass stars and brown dwarfs ($\lesssim 5 \%$), because in most of the cases the dynamical interactions involved in the ejection of the ultracool dwarfs lead to the disruption of the multiple systems. Their important results obtained using high-resolution hydrodynamical simulations of 5 bodies star forming regions \citep{delgado03} and star formation from the fragmentation of turbulent molecular clouds \citep{bate02} are thus in contradiction with the results we present here on that point. Indeed, even if the binary frequency we report here is not much higher than the upper limit predicted by these models, we have to keep in mind that our value is a lower limit since we did not take in account the spectroscopic binaries. Our binary frequency is also in contradiction with the few-body decay models \citep{sterzik98, durisen01} which predict that almost all brown dwarf binaries are broken by gravitational interactions if they originate from bound clusters of moderate sizes. The binary frequency we present here should allow to constrain better the properties of the initial molecular clouds in which ultra-cool dwarfs are supposed to form.

\subsection{The brown dwarf wide binary desert}

The lack of binaries with wide separations that we observe for ultra-cool and brown dwarfs suggest that there are major differences in the formation and evolution processes of these ultra-cool objects in comparison with stars. Whereas they cannot explain the binary frequency we report here, the models of ejection \citep{delgado03, bate02, durisen01} are consistent with the wide binary desert. Only a close binary might indeed be able to survive during the disruption process. As we observe a relatively high rate of binaries, it is more likely that the observed desert might be the consequence of several effects: both formation peculiarities and later evolution processes, such as disruption due to the gravitational interaction with neighbouring stars and/or molecular clouds (see \citet{burgasser03} for a discussion on this last point). These results are consistent with the previous results presented by \citet{reid01, close02a, close02b}.

\subsection{Distribution of Mass Ratios}
The transformation of flux ratio to mass ratio is not straightforward, because the ages of various individual systems can be very different, and second because of the degeneracy in the mass-luminosity (age-temperature) relation in the case of brown dwarfs. Luminosities and effective temperatures of brown dwarfs are indeed function of both age and mass \citep{burrows97, chabrier00}. But since the two components of a binary system can be assumed to be coeval, a difference in the luminosity must be associated to a difference in mass. 

As we observe a strong lack of binaries between $1.5 \lesssim \Delta m_{F814W} \lesssim 2.5$ mag (see Fig. \ref{distri_dmag}), and if we consider that the sample covers a randomly large range of ages (typically between 0.5 Gyr and 10 Gyrs), we can suggest that this might be evidence of a preference for equal-mass systems, as it was thought before. Indeed Table \ref{model} shows that the mass ratios corresponding to $\Delta m_{F814W}=1.5$ mag for ages between 0.5 Gyr and 5.0 Gyrs are ranging between 75$\%$ and 95$\%$, whereas mass ratios corresponding to $\Delta m_{F814W}=4.0$ mag are ranging between 55$\%$ and 74$\%$ \citep{chabrier00}. If we consider a coeval binary system with a primary with $T_{eff}(A)=2012$K, 1Gyr and M$_{A}$=0.07M$_{\sun}$, a difference of magnitude $\Delta m_{F814W}=3$ mag corresponds to a secondary with $T_{eff}(B) \sim 1600$K and M$_{B}$=0.06M$_{\sun}$, whereas a a difference of magnitude $\Delta m_{F814W}=5.5$ mag corresponds to a secondary with $T_{eff}(B) \sim 1500$K and M$_{B}$=0.052M$_{\sun}$. These examples illustrates the limits we were able to reach in our study.

It is also interesting to mention that this lack is similar to the deficiency of low mass ratios among binary early-M dwarfs in comparison with G dwarfs reported by \citet{fischer92} and to the excess of near equal mass M-dwarf binary systems in comparison with G dwarfs systems reported by \citet{reid97}. All these observations suggest different formation mechanisms and/or different processes of evolution. 

The survey is of course biased by the limited sensitivity at smaller separations/lower flux ratios, but nevertheless covers a relatively large dynamic and resolution range. Because of these limits of detection we cannot exclude the possibility of a multi-modal distribution of mass ratio, even if this scenario appears very unlikely in the current context of models of formation and evolution. It is important to remember that we were not able to detect systems with $\Delta m_{F814W}$ greater than 5.5 mag, corresponding to systems with mass ratios of about 45$\%$, 61$\%$, 71$\%$ respectively at the ages of 0.5, 1.0 and 5 Gyrs and for a primary mass of 0.1M$_{\sun}$ (according to the DUSTY models of \citet{chabrier00}, cf. Table \ref{model}).  

This paper has been first submitted in August 2002, but its publication has been delayed untill September 2003. Due to this long delay, several other relevant referenced results have been published since the time of submission.

\section{Summary}
We have reviewed the results of high spatial resolution observations of 134 late M- and L-dwarfs in the field. 26 of them are resolved binaries, with separations between 0.06 and 0.6 arcsec, but one is associated to a star in a triple system. Eleven of the binaries were already known to be multiple systems. Our sample, added to HST data from the archive, makes a sample of 134 objects. The results of these combined observations indicate an binary fraction of $7.6^{+5.9}_{-1.7}\%$ (after correction of the bias) among field ultracool dwarfs with distances less than 25 pc, with all separations below 15 A.U. Considering that our bias correction is over-estimated, the true  HST visual binary fraction should be around 10\%. New statistical studies on better defined samples are necessary to confirm these values. The distribution of flux ratios might be an indication that ultracool dwarf binaries form in nearly equal mass systems ($75\% \lesssim$M2/M1$\lesssim 95 \%$) rather than over the entire range of mass ratios. If confirmed by better statistical studies, these results would be very promising for the future studies: measuring masses of binaries should allow to get the missing constraints for the theoretical models of formation and evolution. Studying the coeval components of multiple systems should allow to calibrate the important mass-luminosity (age-temperature) relation, but also help to ascertain the formation mechanisms. More observations of multiple systems and second and third epoch data will be required in order to constrain the theoretical models and improve our knowledge of very low-mass stars and brown dwarfs.

\acknowledgments
The authors would like to thank Isabelle Baraffe for her useful comments and for providing values of the \citet{chabrier00} and \citet{allard01} DUSTY and NEXTGEN models convolved with the filters of the instruments used for the observations, as well as the anonymous referee of this manuscript for her/his comments and corrections that allow to improve considerably the paper. H. Bouy would like to thank B. Vandame for the very useful discussions and advices on the PSF fitting program and its calibration.

This research is based on observations with the NASA/ESA Hubble Space Telescope, obtained at the Space Telescope Science Institute, and was funded by HST Grants GO8720 and GO9157. This research has made use of data of the NASA/ESA Hubble Space Telescope obtained from the ESO/ST-ECF Science Archive Facility. This publication has made use of the Hipparcos Space Astrometry Mission's results. This publication makes use of data products from the Two Micron All Sky Survey, which is a joint project of the University of Massachusetts and the Infrared Processing and Analysis Center/California Institute of Technology, funded by the National Aeronautics and Space Administration and the National Science Foundation. This publication makes of data from the DEep Near Infrared Survey, and from the Sloan Digital Sky Survey. This research has also made use of Davy Kirkpatrick's on-line archive of L and T Dwarfs \citep{kirk_url}.

\clearpage

\clearpage

\clearpage 

\clearpage

\clearpage 

\clearpage 

\clearpage 

\clearpage

\clearpage

\clearpage 

\clearpage 

\clearpage

\clearpage 

\clearpage

\clearpage
\begin{deluxetable}{lccccccccl}
\tabletypesize{\scriptsize}
\tablecaption{List of Targets. \label{targets}}
\tablewidth{0pt}
\tablehead{
\colhead{Name} & \colhead{R.A \tablenotemark{a}}   & \colhead{Dec \tablenotemark{a}} & \colhead{SpT} & \colhead{I} & \colhead{J}  & \colhead{H} & \colhead{K} &\colhead{Obs. Date} & \colhead{Dist.\tablenotemark{c}} 
}
\startdata
\cutinhead{GO8720}
DENIS-PJ020529.0-115925 \tablenotemark{\star} & 02 05 29.0 & -11 59 25 & L7 & 18.30 & 14.43 & 13.61 & 13.00 & 2000-10-28 &  19.8  \tablenotemark{\bullet}  \\ 
DENIS-PJ024351.0-543219 & 02 43 51.0 & -54 32 19 & M9 & 17.60 & 14.13 & \nodata & 12.93 & 2001-07-02 &  34.4  \\ 
DENIS-P030149-590302 & 03 01 49.0 & -59 03 02 & $\sim$L0 \tablenotemark{b} & 16.84 & 13.71 & \nodata & 12.56 & 2001-03-14 &   31.3  \\ 
DENIS-PJ031433.1-462339 & 03 14 33.1 & -46 23 39 & $\sim$L0  \tablenotemark{b} & 17.99 & 14.88 & \nodata & 13.71 & 2001-06-12 &  54.7  \\ 
DENIS-P035726.9-441730 \tablenotemark{\star \star} & 03 57 26.9 & -44 17 30 & $\sim$L3 \tablenotemark{b} & 18.07 & 14.58 & \nodata & 12.91 & 2001-04-07 &  22.2  \\ 
DENIS-PJ042655.9-573551 & 04 26 55.9 & -57 35 51 & $\sim$L1  \tablenotemark{b} & 18.45 & 15.28 & \nodata & \nodata & 2001-03-03 &  62.3  \\ 
DENIS-P044111.3+010554 & 04 41 11.3 & +01 05 54 & $\sim$L0  \tablenotemark{b} & 19.38 & 16.27 & \nodata & \nodata & 2001-01-14 &  103.7  \\ 
DENIS-PJ090957.1-065806 & 09 09 57.1 & -06 58 06 & L0 & 17.21 & 13.9 & 13.09 & 12.55 & 2000-10-29 &  26.4  \\ 
DENIS-P100428.3-114648 \tablenotemark{\star \star} & 10 04 28.3 & -11 46 48 & $\sim$L0  \tablenotemark{b} & 18.0 & 14.9 & \nodata & 13.67 & 2000-10-27 &  46.8  \\ 
DENIS-P101621.9-271428 & 10 16 21.9 & -27 14 28 & $\sim$L2 \tablenotemark{b} & 18.45 & 15.10 & \nodata & \nodata & 2000-10-22 &  47.1  \\ 
DENIS-P104731.1-181558 & 10 47 31.1 & -18 15 58 & L2.5 & 17.75 & 14.24 & \nodata & 12.88 & 2001-07-27 &  20.9  \\ 
DENIS-P104814.7-395606 & 10 48 14.7 & -39 56 06 & $\sim$L0  \tablenotemark{b} & 12.67 & 9.59 & \nodata & 8.58 & 2001-04-07 &  4.9  \\ 
2MASSWJ1145572+231730 & 11 45 57.2 & +23 17 30 & L1.5 & 18.62 & 15.37 & 14.52 & 13.92 & 2001-04-04 &  41.1  \\ 
DENIS-PJ115442.2-340039 & 11 54 42.2 & -34 00 39 & $\sim$L4 \tablenotemark{b} & 17.90 & 14.26 & \nodata & 12.64 & 2001-04-07 &  20.9  \\ 
DENIS-P121612.1-125731 & 12 16 12.1 & -12 57 31 & $\sim$L1 \tablenotemark{b} & 18.30 & 15.11 & \nodata & \nodata & 2001-03-04 &  56.5  \\ 
DENIS-PJ122813.8-154711 \tablenotemark{\star} & 12 28 15.2 & -15 47 34 & L5 & 18.19 & 14.43 & \nodata & 12.73 & 2001-03-04 &  20.2  \tablenotemark{\bullet}   \\ 
DENIS-PJ122821.6-241541 & 12 28 21.6 & -24 15 41 & $\sim$L5  \tablenotemark{b} & 18.00 & 14.28 & 13.4 & 12.71 & 2001-02-27 &  20.2  \\ 
DENIS-PJ131500.9-251302 & 13 15 00.9 & -25 13 02 & $\sim$M9  \tablenotemark{b} & 18.21 & 15.16 & \nodata & \nodata & 2001-03-29 &  65.3  \\ 
2MASSW1342236+175156  & 13 42 23.6 & +17 51 56 & L2.5 & 19.81 & 16.06 & 15.12 & 14.59 & 2000-11-23 &  48.2  \\ 
2MASSIJ1346464-003150 & 13 46 46.4 & -00 31 50 & T & 20.0 & 15.86 & 16.05 & 15.74 & 2001-03-15 &  15.5  \\ 
DENIS-PJ141217.1-043358 & 14 12 17.1 & -04 33 58 & $\sim$L5 \tablenotemark{b} & 18.78 & 15.05 & \nodata & 13.61 & 2001-03-27 &  25.7  \\ 
2MASSWJ1439409+182637 & 14 39 40.9 & +18 26 37 & L1 & 19.87 & 16.21 & 15.47 & 14.53 &  2001-03-26 &  65.4  \\ 
SDSS144001.8+002145.8 & 14 40 01.8 & +00 21 45.8 & L1 & 18.8 & 15.9 & 15.1 & 14.6 & 2001-01-28 &  101.1  \\ 
DENIS-PJ144137.3-094559 \tablenotemark{\star} & 14 41 37.3 & -09 45 59 & L1 & 17.32 & 14.25 & \nodata & 12.37 & 2001-01-16 &  29.2  \\ 
DENIS-PJ161928.3+005012 & 16 19 28.3 & +00 50 12 & $\sim$L2 \tablenotemark{b} & 17.79 & 14.40 & \nodata & 13.01 & 2001-06-02 &  32.4  \\ 
DENIS-PJ191903.9-413433 & 19 19 03.9 & -41 34 33 & $\sim$L3 \tablenotemark{b} & 19.4 & 15.9 & \nodata & \nodata & 2000-08-03 &  55.6  \\ 
DENIS-PJ202333.6-181500 & 20 23 33.6 & -18 15 00 & $\sim$L3 \tablenotemark{b} & 18.50 & 15.0 & \nodata & \nodata & 2001-03-24 &  36.7  \\ 
DENIS-PJ221538.2-080912 & 22 15 38.2 & -08 09 12 & $\sim$L1 \tablenotemark{b} & 17.86 & 14.56 & \nodata & 13.26 & 2001-06-28 &  39.0  \\ 
DENIS-PJ232931.7-540858 & 23 29 31.7 & -54 08 58 & M8  \tablenotemark{b} & 18.18 & 15.15 & \nodata & 13.53 & 2001-07-03 &  64.3  \\ 

\cutinhead{GO8146}

2MASSW003616+182110 & 00 36 16.0 & +18 21 10 & L3.5 & 16.0 & 12.44 & 11.58 & 11.03 &    2000-02-15 &  08.7  \tablenotemark{\bullet} \\ 
2MASSW0708213+295035 & 07 08 21.3 & +29 50 35 & L5  & 20.2 & 16.75 & 15.54 & 14.69 &    2000-03-23 &  44.8  \\ 
2MASSW0740096+321203 & 07 40 09.6 & +32 12 03 & L4.5 & 19.6 & 16.2 & 14.8 & 14.2 &      2000-03-27 &  37.6  \\ 
2MASSW0746425+200032 \tablenotemark{\star}& 07 46 42.5 & +20 00 32 & L0.5 & 15.1 & 11.7 & 11.0 & 10.5 & 2000-04-15 &  12.3  \tablenotemark{\bullet} \\ 
2MASSW0820299+450031 & 08 20 29.9 & +45 00 31 & L5 & 20.0 & 16.3 & 15.0 & 14.2 &        2000-04-24 &  36.4  \\ 
2MASSW0825196+211552 & 08 25 19.6 & +21 15 52 & L7.5 & 18.9 & 15.1 & 13.8 & 13.0 &      2000-03-25 &  10.7  \tablenotemark{\bullet} \\ 
2MASSW0850359+105715 \tablenotemark{\star}& 08 50 35.9 & +10 57 15 & L6 & \nodata & 16.5 & 15.2 & 14.5 &   2000-02-01 &  41.0  \tablenotemark{\bullet} \\ 
2MASSW0913032+184150 & 09 13 03.2 & +18 41 50 & L3 & 19.4 & 15.9 & 14.8 & 14.2 &        2000-04-05 &  41.4  \\ 
2MASSW0920122+351742 \tablenotemark{\star}& 09 20 12.2 & +35 17 42 & L6.5 & 19.4 & 15.6 & 14.7 & 13.9 & 2000-02-09 & 20.1  \\ 
2MASSW0928397-160312 & 09 28 39.7 & -16 03 12 & L2 & 18.8 & 15.3 & 14.3 & 13.6 &        2000-04-28 &  36.8  \\ 
2MASSW1029216+162652 & 10 29 21.6 & +16 26 52 & L2.5 & \nodata & 14.3 & 13.3 & 12.6 &      2000-03-06 &  21.4  \\ 
2MASSW1123556+412228 & 11 23 55.6 & +41 22 28 & L2.5 & 19.6 & 16.1 & 15.1 & 14.3 &      2000-04-19 &  21.7  \tablenotemark{\bullet}   \\ 
2MASSW1146344+223052 \tablenotemark{\star}& 11 46 34.4 & +22 30 52 & L3 & \nodata & 14.2 & 13.2 & 12.6 &   2000-04-28 &  27.2  \tablenotemark{\bullet}  \\ 
2MASSW1155008+230705 & 11 55 00.8 & +23 07 05 & L4 & 19.5 & 15.8 & 14.7 & 14.1 &        2000-03-18 &  33.8  \\ 
2MASSW132855+211449 & 13 28 55.0 & +21 14 49 & L5 & 19.8 & 16.1 & 16.0 & 14.3 &        2000-04-23 &  32.2  \tablenotemark{\bullet} \\ 
2MASSW1338261+414034 & 13 38 26.1 & +41 40 34 & L2.5 & 17.6 & 14.2 & 13.3 & 12.8 &      2000-04-25 &  20.5  \\ 
2MASSW1343167+394508 & 13 43 16.7 & +39 45 08 & L5 & 19.8 & 16.2 & 14.9 & 14.1 &        2000-04-21 &  34.7  \\ 
2MASSW1439284+192915 & 14 39 28.4 & +19 29 15 & L1 & 16.1 & 12.8 & 12.0  & 11.6 &       2000-03-22 &  14.4  \tablenotemark{\bullet} \\ 
2MASSW1507476-162738 & 15 07 47.6 & -16 27 38 & L5 & 17.0 & 12.8 & 11.9 & 11.3 &        2000-02-24 &  07.3  \tablenotemark{\bullet}   \\ 
2MASSW1632291+190440 & 16 32 29.1 & +19 04 40 & L8 & 19.7 & 15.9 & 14.6 & 14.0 &        2000-04-20 &  15.2  \tablenotemark{\bullet} \\ 
2MASSW1726000+153819 & 17 26 00.0 & +15 38 19 & L2 & 19.4 & 15.7 & 14.5 & 13.6 &        2000-03-24 &  44.2  \\ 

\cutinhead{GO8581}
2MASSW0010036+343609 & 00 10 03.6 & +34 36 09 & M8 & 18.5 & 15.6 & 15.1 & 14.4 & 2001-01-20 & 79.1 \\
SDSS0019117+0030179 & 00 19 11.7 & +00 30 17.9 & L1 & 18.4 & 14.9 & 14.2 & 13.6 & 2001-07-09 & 35.8 \\
2MASSW0028394+150141 & 00 28 39.4 & +15 01 41 & L4.5 & 20.0 & 16.5 & 15.3 & 14.6 & 2001-01-14 & 43.2 \\
2MASSW0030300-145033 & 00 30 30.0 & -14 50 33 & L7 & 18.6 & 16.8 & 15.4 & 14.4 & 2001-01-13 & 33.5 \\
2MASSW0033239-152131 & 00 33 23.9 & -15 21 30.9 & \nodata & 18.3 & \nodata & \nodata & \nodata & 2001-05-13 & \nodata \\
2MASSW0208183+254253 & 02 08 18.3 & +25 42 53 & L1 & 17.7 & 14.0 & 13.1 & 12.6 & 2000-08-26 & 23.6 \\
2MASSW0224366+253704 & 02 24 36.6 & +25 37 04 & L2 & 19.7 & 16.6 & 15.4 & 14.7 & 2001-01-09 & 66.9 \\
2MASSW0326422-210205 & 03 26 42.2 & -21 02 05 & L3.5\tablenotemark{b} & 19.9 & 16.11 & 14.77 & 13.88 & 2000-12-30 & 32.2 \tablenotemark{\bullet} \\
2MASSW0328426+230205 & 03 28 42.6 & +23 02 05 & L8 & 20.3 & 16.7 & 15.6 & 14.8 & 2001-02-28 & 27.3 \\
SDSS0330351-002534 & 03 30 35.1 & -00 25 34 & L2 & 19.0 & 15.29 & 14.42 & 13.83 & 2001-06-30 & 36.6 \\
2MASSW0335020+234235 & 03 35 02.0 & +23 42 35 & $\sim$L2\tablenotemark{b} & 15.7 & 12.26 & 11.65 & 11.26 & 2001-07-21 & 11.3 \\
2MASSW0337036-175807 & 03 37 03.6 & -17 58 07 & L4.5 & 19.4 & 15.6 & 14.4 & 13.6 & 2000-12-25 & 28.5 \\
SDSS0344089+011125 & 03 44 08.9 & +01 11 25.0 & \nodata & 18.2 & \nodata & \nodata & \nodata & 2001-07-12 & \nodata \\
2MASSW0345432+254023 & 03 45 43.2 & +25 40 23 & L0 & 17.7 & 14.0 & 13.2 & 12.7 & 2001-03-27 & 26.9 \tablenotemark{\bullet} \\
2MASSW0350573+181806 & 03 50 57.3 & +18 18 06 & $\sim$L2\tablenotemark{b} & 16.4 & 13.0 & 12.2 & 11.8 & 2001-02-26 & 16.8 \\
2MASSW0355419+225701 & 03 55 41.9 & +22 57 01 & L3 & 19.6 & 16.1 & 15.0 & 14.2 & 2000-08-25 & 45.4 \\
SDSS0539519-005901 & 05 39 51.9 & -00 59 01 & L5 & 18.0 & 14.02 & 13.09 & 12.51 & 2000-11-27 & 12.7 \\
2MASSW0753321+291711 & 07 53 32.1 & +29 17 11 & L2 & 19.0 & 15.5 & 14.5 & 13.8 & 2000-09-08 & 40.3 \\
2MASSW0801405+462850 & 08 01 40.5 & +46 28 50 & L6.5 & 19.7 & 16.3 & 15.4 & 14.5 & 2001-03-08 & 28.7 \\
2MASSW0829570+265510 & 08 29 57.0 & +26 55 10 & L6.5 & 20.6 & 17.0 & 15.8 & 14.9 & 2001-04-18 & 39.7 \\
2MASSW0832045-012835 & 08 32 04.5 & -01 28 35 & L1.5 & 17.9 & 14.1 & 13.3 & 12.7 & 2000-09-23 & 22.9 \\
2MASSW0856479+223518\tablenotemark{\star\star} & 08 56 47.9 & +22 35 18 & L3 & 19.2 & 15.65 & 14.58 & 13.92 & 2001-04-24 & 34.7 \\
2MASSW0914188+223813 & 09 14 18.8 & +22 38 13 & $\sim$L1\tablenotemark{b} & 18.6 & 15.30 & 14.40 & 13.90 & 2001-04-07 & 54.8 \\
2MASSW0951054+355802 & 09 51 05.4 & +35 58 02 & L6 & 20.7 & 17.1 & 15.8 & 15.1 & 2001-01-26 & 44.9 \\
2MASSW1017075+130839\tablenotemark{\star\star} & 10 17 07.5 & +13 08 39 & L3 & 17.8 & 14.1 & 13.2 & 12.7 & 2001-04-16 & 21.4 \\
SDSS1043251+000148 & 10 43 25.1 & +00 01 48 & L3 & 19.3 & 15.95 & 15.17 & 14.53 & 2001-05-02 & 42.4 \\
2MASSW1102337-235945 & 11 02 33.7 & -23 59 45 & L4.5 & 20.4 & 17.0 & 15.6 & 14.8 & 2001-05-10 & 54.3 \\
2MASSW1104012+195921 & 11 04 01.2 & +19 59 21 & L4.5 & 18.3 & 14.4 & 13.5 & 13.0 & 2001-04-26 & 16.4 \\
2MASSW1108307+683017 & 11 08 30.7 & +68 30 17 & L1 & 17.2 & 13.31 & 12.2 & 11.92 & 2000-10-02 & 17.2 \\
2MASSW1112256+354813\tablenotemark{\star\star} & 11 12 25.6 & +35 48 13 & L4.5 & 18.5 & 14.6 & 13.5 & 12.7 & 2001-02-14 & 21.7 \tablenotemark{\bullet} \\
2MASSW1127534+741107\tablenotemark{\star\star} & 11 27 53.4 & +74 11 07 & $\sim$L2\tablenotemark{b} & 16.5 & 13.1 & 12.4 & 12.0 & 2001-05-05 & 14.6 \\
2MASSW1239194+202952 & 12 39 19.4 & +20 29 52.0 & \nodata & 17.8 & \nodata & \nodata & \nodata & 2001-05-13 & \nodata \\
2MASSW1239272+551537\tablenotemark{\star\star} & 12 39 27.2 & +55 15 37 & L5 & 18.6 & 14.7 & 13.5 & 12.7 & 2001-03-18 & 21.3 \\
2MASSW1311391+803222\tablenotemark{\star\star} & 13 11 39.1 & +80 32 22 & $\sim$L2\tablenotemark{b} & 16.2 & 12.8 & 12.1 & 11.7 & 2000-07-30 & 13.7 \\
2MASSW1403223+300754 & 14 03 22.3 & +30 07 54 & M8.5 & 16.3 & 12.7 & 12.0 & 11.6 & 2001-06-14 & 19.2 \\
2MASSW1411175+393636 & 14 11 17.5 & +39 36 36 & L1.5 & 18.2 & 14.7 & 13.8 & 13.3 & 2000-09-19 & 30.2 \\
2MASSW1412244+163312 & 14 12 24.4 & +16 33 12 & L0.5 & 17.6 & 13.89 & 13.06 & 12.59 & 2000-09-02 & 24.3 \\
2MASSW1426316+155701\tablenotemark{\star} & 14 26 31.6 & +15 57 01.3 & M9 & 16.5 & 12.87 & 12.18 & 11.71 & 2001-07-19 & 26.7 \\
2MASSW1430435+291540\tablenotemark{\star\star} & 14 30 43.5 & +29 15 40 & L2 & 18.0 & 14.3 & 13.4 & 12.7 & 2001-04-19 & 29.4 \\
2MASSW1434264+194050 & 14 34 26.4 & +19 40 50.0 & \nodata & 18.5 & \nodata & \nodata & \nodata & 2001-05-11 & \nodata\\ 
SDSS1435172-0046129 & 14 35 17.2 & -00 46 12.9 & L0 & 19.7 & 16.5 & 15.6 & 15.3 & 2001-07-23 & 87.5 \\
SDSS1435357-004347 & 14 35 35.7 & -00 43 47 & L3 & 19.2 & 16.5 & 15.7 & 15.0 & 2001-05-11 & 54.6 \\
2MASSW1438549-130910 & 14 38 54.9 & -13 09 10 & L3 & 19.1 & 15.5 & 14.5 & 13.9 & 2001-03-12 & 34.5 \\
2MASSW1438082+640836 & 14 38 08.2 & +64 08 36 & $\sim$L6\tablenotemark{b} & 16.8 & 12.92 & 12.03 & 11.57 & 2001-03-10 & 7.2 \\
2MASSW1449378+235537\tablenotemark{\star\star} & 14 49 37.8 & +23 55 37 & L0 & 18.9 & 15.6 & 15.0 & 14.5 & 2000-12-21 & 63.7 \\
2MASSW1457396+451716 & 14 57 39.6 & +45 17 16 & $\sim$L3\tablenotemark{b} & 16.7 & 13.1 & 12.4 & 11.9 & 2000-10-31 & 13.1 \\
2MASSW1506544+132106 & 15 06 54.4 & +13 21 06 & L3 & 17.4 & 13.4 & 12.4 & 11.7 & 2001-05-09 & 13.1 \\
SDSS151547.2-0030597 & 15 15 47.2 & -00 30 59.7 & \nodata & 17.5 & \nodata & \nodata & \nodata & 2001-05-06 & \nodata \\
2MASSW1526140+204341 & 15 26 14.0 & +20 43 41 & L7 & 19.1 & 15.6 & 14.5 & 13.9 & 2001-05-09 & 19.2 \\
SDSS154831.7+0029415 & 15 48 31.7 & -00 29 41.5 & \nodata & 18.5 & \nodata & \nodata & \nodata & 2001-05-30 & \nodata\\ 
2MASSW1550382+3041037 & 15 50 38.2 & +30 41 03.7 & \nodata & 16.3 & \nodata & \nodata & \nodata & 2001-05-10 & \nodata \\
2MASSW1551066+645704 & 15 51 06.6 & +64 57 04 & $\sim$L3\tablenotemark{b} & 16.5 & 12.9 & 12.1 & 11.7 & 2000-09-17 & 12.0 \\
2MASSW1600054+170832\tablenotemark{\star\star} & 16 00 05.4 & +17 08 32 & L1.5 & 19.3 & 16.1 & 15.1 & 14.7 & 2001-01-14 & 60.6 \\
2MASSW1627279+810507 & 16 27 27.9 & +81 05 07 & $\sim$L5\tablenotemark{b} & 16.8 & 13.0 & 12.3 & 11.9 & 2001-03-15 & 19.0 \\
2MASSW1635191+422305 & 16 35 19.1 & +42 23 05 & $\sim$L3\tablenotemark{b} & 16.5 & 12.9 & 12.2 & 11.8 & 2001-05-12 & 12.0 \\
SDSS1653297+6231365 & 16 53 29.7 & +62 31 36.5 & L3 & 18.1 & 15.1 & 14.4 & 13.9 & 2001-05-04 & 28.7 \\
2MASSW1656188+283506 & 16 56 18.8 & +28 35 06 & L4.5 & 20.3 & 17.1 & 15.9 & 15.0 & 2001-04-25 & 56.9 \\
2MASSW1707333+430130 & 17 07 33.3 & +43 01 30 & L0 & 17.6 & 14.0 & 13.2 & 12.7 & 2001-05-13 & 27.7 \\
2MASSW1707183+6439334 & 17 07 18.3 & +64 39 33.4 & \nodata & 16.3 & \nodata & \nodata & \nodata & 2001-05-13 & \nodata \\
2MASSW1710254+210715 & 17 10 25.4 & +21 07 15 & $\sim$M8\tablenotemark{b} & 18.6 & 15.87 & 15.02 & 14.46 & 2001-03-19 & 106.7 \\
2MASSW1711457+223204 & 17 11 45.7 & +22 32 04 & L6.5 & 20.4 & 17.1 & 15.8 & 14.7 & 2001-03-20 & 41.6 \\
SDSS1723287+6406233 & 17 23 28.7 & +64 06 23.3 & \nodata & 19.1 & \nodata & \nodata & \nodata & 2001-05-30 & \nodata\\ 
2MASSW1728114+394859\tablenotemark{\star\star} & 17 28 11.4 & +39 48 59 & L7 & 19.6 & 16.0 & 14.8 & 13.9 & 2000-08-12 & 20.4 \\
2MASSW1743349+212711 & 17 43 34.9 & +21 27 11 & L2.5 & 19.3 & 15.80 & 14.78 & 14.29 & 2001-01-24 & 42.8 \\
2MASSW1743348+584411 & 17 43 34.8 & +58 44 11 & L0.5 & 17.7 & 14.02 & 13.15 & 12.67 & 2001-06-15 & 25.8 \\
2MASSW1841086+311727 & 18 41 08.6 & +31 17 27 & L4 & 19.6 & 16.1 & 15.0 & 14.2 & 2001-03-08 & 38.8 \\
2MASSW2054358+151904 & 20 54 35.8 & +15 19 04 & L1 & 19.6 & 16.5 & 15.6 & 14.8 & 2001-05-03 & 74.8 \\
2MASSW2057153+171515 & 20 57 15.3 & +17 15 15 & L1.5 & 19.4 & 16.1 & 15.2 & 14.6 & 2001-05-02 & 57.5 \\
2MASSW2101154+175658\tablenotemark{\star\star} & 21 01 15.4 & +17 56 58 & L7.5 & 20.1 & 16.87 & 15.70 & 15.04 & 2001-05-07 & 23.2 \\
2MASSW2140293+162518\tablenotemark{\star} & 21 40 29.3 & +16 25 18 & $\sim$L2\tablenotemark{b} & 16.4 & 12.9 & 12.3 & 11.8 & 2001-05-31 & 12.9 \\
2MASSW2147436+143131\tablenotemark{\star\star} & 21 47 43.6 & +14 31 31 & $\sim$L2\tablenotemark{b} & 17.3 & 13.8 & 13.1 & 12.7 & 2000-10-09 & 21.8 \\
2MASSW2158045-155009 & 21 58 04.5 & -15 50 09 & L4 & 18.8 & 14.9 & 13.9 & 13.1 & 2000-10-30 & 22.3 \\
2MASSW2206449-421720 & 22 06 44.9 & -42 17 20 & L2 & 19.2 & 15.6 & 14.5 & 13.6 & 2001-03-22 & 42.2 \\
2MASSW2206228-204705\tablenotemark{\star} & 22 06 22.8 & -20 47 05 & M8 & 16.0 & 12.4 & 11.7 & 11.3 & 2000-08-13 & 22.2 \tablenotemark{\bullet}\\
2MASSW2208136+292121 & 22 08 13.6 & +29 21 21 & L2 & 19.5 & 15.82 & 14.83 & 14.09 & 2000-10-12 & 46.7 \\
2MASSW2234139+235955 & 22 34 13.9 & +23 59 55 & M9.5 & 16.5 & 13.2 & 12.4 & 11.8 & 2001-01-14 & 20.7 \\ 
2MASSW2242531+254257 & 22 42 53.1 & +25 42 57 & $<$M5\tablenotemark{b} & 18.3 & 14.8 & 13.8 & 13.0 & 2001-04-26 & \nodata\\ 
2MASSW2244316+204343 & 22 44 31.6 & +20 43 43 & L6.5 & 20.1 & 16.53 & 14.97 & 13.97 & 2001-05-09 & 11.3 \tablenotemark{\bullet} \\
2MASSW2306292-050227 & 23 06 29.2 & -05 02 27 & $\sim$L4\tablenotemark{b} & 15.1 & 11.37 & 10.72 & 10.29 & 2000-08-18 & 8.6 \\
2MASSW2309462+154905 & 23 09 46.2 & +15 49 05 & \nodata & 18.2 & \nodata & \nodata & \nodata & 2000-10-10 & \nodata \\
2MASSW2331016-040619\tablenotemark{\star} & 23 31 01.6 & -04 06 19 & $\sim$L2\tablenotemark{b} & 16.3 & 12.9 & 12.3 & 11.9 & 2001-05-06 & 26.2 \tablenotemark{\bullet} \\
SDSS2335583-001304\tablenotemark{\star\star} & 23 35 58.3 & -00 13 04.0 & \nodata & 18.9 & \nodata & \nodata & \nodata & 2001-06-29 & \nodata \\
2MASSW2349489+122438 & 23 49 48.9 & +12 24 38 & $\sim$L4\tablenotemark{b} & 16.2 & 12.6 & 12.0 & 11.6 & 2001-05-13 & 19.2 \\

 & & & & & & & & &  \\ 

\enddata
\tablenotetext{\star}{ indicates the previously known binaries}
\tablenotetext{\star \star}{ indicates the new multiple systems candidates presented in this paper} 
\tablenotetext{a}{ J2000}
\tablenotetext{b}{ Spectral Types estimated using the SpT vs. (I-J) relation given in \citet{dahn2002}}
\tablenotetext{c}{ Distances in pc estimated as explained in the text or by trigonometric parallax \citep{dahn2002, perryman97} when available (these latter cases are indicated with a sign $\bullet$)}
\end{deluxetable}

\clearpage

\begin{deluxetable}{lccccc}
\tabletypesize{\scriptsize}
\tablecolumns{6}
\tablewidth{0pc}
\tablecaption{Systematic errors and 1$\sigma$ uncertainties on the photometric results ($\Delta m_{F814W}$)\label{table_calib1}}
\tablehead{
\colhead{} & \multicolumn{2}{c}{$\delta <$ 0\farcs150} &  \colhead{} & \multicolumn{2}{c}{$\delta >$ 0\farcs150}  \\
\cline{2-3} \cline{5-6}
\colhead{$\Delta m_{F814W}$} & \colhead{Systematic error\tablenotemark{a}}  & \colhead{1$\sigma$}  & \colhead{} & \colhead{Systematic error} & \colhead{1$\sigma$}
}

\startdata
0.00 & $-6.9+171.4\times\delta-1425.3\times\delta^{2}+3948.7\times\delta^{3}$ & $\pm$0.05 &  & -0.01 & $\pm$0.01 \\
0.11 & $-5.6+132.4\times\delta-1033.8\times\delta^{2}+2674.5\times\delta^{3}$ & $\pm$0.05 &  & 0.09 & $\pm$0.02 \\
0.11 $\to$ 0.24 & $-5.3+130.5\times\delta-1036.8\times\delta^{2}+2713.5\times\delta^{3}$ & $\pm$0.07 &  & 0.17 & $\pm$0.07 \\
0.24 $\to$ 2.50 & $-3.5+103.3\times\delta - 925.4\times\delta^{2}+2687.6\times\delta^{3}$ & $\pm$0.11 &  & 0.20 & $\pm$0.09 \\
\enddata
\tablenotetext{a}{The systematic errors (in mag) are obtained using the given relation, with $\delta$ corresponding to the separation in arcseconds.}
\tablecomments{The systematic errors given in this table had to be added to the value we measured. See also Figure \ref{calib3}.}

\end{deluxetable}

\clearpage

\begin{deluxetable}{lll}
\tablewidth{0pc}
\tablecaption{Systematic errors and 1$\sigma$ uncertainties on the astrometric results\label{table_calib2}}
\tablehead{
\colhead{} & \colhead{$\delta <$ 0\farcs150} & \colhead{$\delta >$ 0\farcs150}
}

\startdata
P.A & $\approx0.0^{\circ}\pm1.2^{\circ}$ & $\approx0.0^{\circ}\pm0.3^{\circ}$ \\
Separation & $\approx0\farcs0\pm0\farcs0028$ & $\approx0\farcs0\pm0\farcs0028$ \\ 
\enddata
\tablecomments{see also Figure \ref{calib2} and Table \ref{table_calib1}. $\delta$ represents the separation.}
\end{deluxetable}

\clearpage

\begin{center}
\begin{deluxetable}{lccllll}
\tabletypesize{\scriptsize}
\tablecaption{Results \label{results}}
\tablewidth{0pt}
\tablehead{
\colhead{Name} & \colhead{Date of Obs.}  & \colhead{Sep. (mas) \tablenotemark{d}} &\colhead{P.A ($^{\circ}$) \tablenotemark{d}}  & \colhead{$\Delta$mag\tablenotemark{d}} & \colhead{mag(A)\tablenotemark{e}} & \colhead{Filter}   
}
\startdata
\cutinhead{GO8720}
DENISPJ0205-1159 & 2000-10-28 & 409 \tablenotemark{b} & 259.7 \tablenotemark{b} & 1.47$\pm$0.19 & 20.75$\pm$0.26 &F675W        \\
 &  2000-10-28 & 398 \tablenotemark{b} & 257.9 \tablenotemark{b} & 0.65$\pm$0.16 & 18.32$\pm$0.24 & F814W \\
 &  2001-07-08  & 370.0  &     255.8 $\pm$     0.3 & 0.63$\pm$0.09 & 18.58$\pm$0.15 & F814W \\
 &  & &  &  \\

DENISPJ0357-4417 & 2001-04-07 & 97.0  & 174 $\pm$ 1.2 & 1.23$\pm$0.11 & 20.37$\pm$0.18 & F675W \\
                 & 2001-04-07 & 98.1 & 174.7$\pm$ 1.2 & 1.50$\pm$0.11 & 17.94$\pm$0.18 & F814W        \\

 & & & &   \\

DENISPJ1004-1146  & 2000-10-27 & 146.0  & 306.1 $\pm$ 1.2 & 0.25$\pm$0.07 & 20.88$\pm$0.18  & F675W\\
 & 2000-10-27 & 146.0 & 304.5 $\pm$1.2 & 0.66$\pm$0.11 & 18.40$\pm$0.18 & F814W\\

 &  & & & \\

DENISPS1228-1547  & 2001-03-04 & 264 \tablenotemark{a}& 21.5 $\pm$ 1.0 & 0.53$\pm$0.09 & 20.52$\pm$0.15 & F675W \\ 
& 2001-03-04 & 246 \tablenotemark{a}& 23.1 $\pm$ 2.0 & 0.44$\pm$0.09 & 18.37$\pm$0.15 & F814W\\
& 2001-06-16 & 255.4  & 18.3 $\pm$ 0.3 & 0.36$\pm$0.07 & 18.34$\pm$0.14 & F814W\\

 &  & & &   \\

DENISPJ1441-0945  & 2001-01-16 & 374.0  & 290.4 $\pm$ 0.3 & 0.63$\pm$0.09 & 20.05$\pm$0.15 & F675W \\   
 & 2001-01-16 & 375.1 & 290.3$\pm$ 0.3 & 0.37$\pm$0.07 & 17.84$\pm$0.14 & F814W \\
 & 2001-05-22 & 370.2 & 290.8$\pm$ 0.3 & 0.30$\pm$0.07 & 17.82$\pm$0.14 & F814W \\

\cutinhead{GO8146}

2MASSW0746+2000 & 2000-04-15 & 219.0 & 168.8 $\pm$ 0.3 & 1.0$\pm$0.09 & 15.41$\pm$0.15 & F814W\\     

 &  & & &    \\

2MASSW0850+1057 & 2000-02-01 & 157.2 & 114.7 $\pm$ 0.3 & 1.47$\pm$0.09 & 20.29$\pm$0.15 & F814W \\     

 &  & & &   \\

2MASSW0920+3517  & 2000-02-09 & 75.1 & 248.5 $\pm$ 1.2 & 0.88$\pm$0.11 & 19.83$\pm$0.18 & F814W \\
        
 &  & & &    \\

2MASSW1146+2230 & 2000-04-28 & 294.1 & 199.5 $\pm$ 0.3 & 0.75$\pm$0.09 & 18.17$\pm$0.15 & F814W \\

\cutinhead{GO8581}
2MASSW0856+2235 & 2001-04-24 & 98 $\pm$ 9 & 275 $\pm$ 2.0  & 2.76$\pm$0.11 & 19.26$\pm$0.20 & F814W \\ 
 &  & & &    \\
2MASSW1017+1308 & 2001-04-16 & 102.0 & 88.2 $\pm$ 1.2 & 0.77$\pm$0.11 & 17.88$\pm$0.18 & F814W \\
                & 2001-04-16 & 104.0 & 92.6 $\pm$ 1.2 & 0.74$\pm$0.11 & 14.82$\pm$0.18 & F1042M \\

 &  & & &   \\

2MASSW1112+3548 & 2001-02-14 & 70.0  & 79.6 $\pm$ 1.2 & 1.07$\pm$0.11 & 18.58$\pm$0.18 & F814W \\
        & 2001-02-14 & 70.0  & 79.2 $\pm$ 1.2 & 1.04$\pm$0.11 & 15.16$\pm$0.18 & F1042M \\
 &  & & &   \\

2MASSW1127+7411 & 2001-05-05 & 252.3  & 80.5 $\pm$ 0.3 & 0.82$\pm$0.09 & 16.25$\pm$0.14 & F814W \\
                & 2001-05-05 & 252.5  & 79.9 $\pm$ 0.3 & 0.39$\pm$0.07 & 14.34$\pm$0.14 & F1042M \\

 &  & & &   \\

2MASSW1239+5515 & 2001-03-18 & 211.0   & 187.6 $\pm$ 0.3 & 0.34$\pm$0.07 & 18.86$\pm$0.15  & F814W \\
                & 2001-03-18 & 210.1 & 184.2 $\pm$ 0.3 & 0.54$\pm$0.09 & 14.62$\pm$0.15 & F1042M \\

 &  & & &   \\

2MASSW1311+8032 & 2000-07-30 & 300.8 & 167.2 $\pm$ 0.3 & 0.39$\pm$0.07 & 16.00$\pm$0.17 & F814W \\
                & 2000-07-30 & 300.0 & 167.3 $\pm$ 0.3 & 0.45$\pm$0.09 & 13.65$\pm$0.17 & F1042M \\
 &  & & &   \\

2MASSW1426+1557 & 2001-07-19 & 157.1 & 339.9 $\pm$ 0.3 & 1.40$\pm$0.09 & 16.15$\pm$0.17 & F814W \\
                & 2001-07-19 & 154.2 & 340.1 $\pm$ 0.3 & 1.30$\pm$0.09 & 13.99$\pm$0.17 & F1042M \\

 &  & & &   \\

2MASSW1430+2915 & 2001-04-19 & 83.0 & 327.1 $\pm$ 1.2 & 0.78$\pm$0.11 & 18.16$\pm$0.18 & F814W \\
                & 2001-04-19 & 88.0 & 327.5 $\pm$ 1.2 & 0.76$\pm$0.11 & 15.13$\pm$0.18 & F1042M \\

 &  & & &   \\

2MASSW1449+2355 & 2000-12-21 & 134.0 & 64.4 $\pm$ 1.2 & 1.51$\pm$0.11 & 18.93$\pm$0.18  & F814W \\
                & 2000-12-21 & 129.0 & 63.4 $\pm$ 1.2 & 1.08$\pm$0.11 & 15.17$\pm$0.18 & F1042M \\
 &  & & &   \\

2MASSW1600+1708 & 2001-01-14 & 57.0 & 156.2 $\pm$ 1.2 & 0.69$\pm$0.11 & 19.81$\pm$0.17 & F814W \\

 &  & & &   \\

2MASSW1728+3948 & 2000-08-12 & 131.3 & 27.6 $\pm$ 1.2 & 0.66$\pm$0.11 & 20.25$\pm$0.17 & F814W \\
 &  & & &   \\

2MASSW2101+1756 & 2001-05-07 & 234.3 & 107.7 $\pm$ 0.3 & 0.59$\pm$0.09 & 20.84$\pm$0.17 & F814W \\

 &  & & &   \\

2MASSW2140+1625 & 2001-05-31 & 159.0 & 132.4 $\pm$ 0.3 & 1.51$\pm$0.09 & 16.04$\pm$0.17 & F814W \\       
 & 2000-05-31 & 157 & 132.8 $\pm$ 0.3 & 1.38$\pm$0.09 & 13.78$\pm$0.17 & F1042M\\

 &  & & &   \\

2MASSW2147+1431 & 2000-10-09 & 322.7 & 329.5 $\pm$ 0.3 & 0.62$\pm$0.09 & 17.38$\pm$0.17 & F814W \\
                & 2000-10-09 & 333\tablenotemark{c} & 332\tablenotemark{c} & 0.75$\pm$0.28\tablenotemark{c} & 14.67$\pm$0.30 & F1042M \\

 &  & & &   \\

2MASSW2206-2047  & 2000-08-13 & 163.0 & 57.5 $\pm$ 0.3 & 0.36$\pm$0.07 & 15.69$\pm$0.14  & F814W \\
 & 2000-08-13 & 160.7 & 57.2 $\pm$ 0.3 & 0.30$\pm$0.07 & 13.84$\pm$0.14 & F1042M\\
 &  & & &    \\

2MASSW2331-0406 & 2001-05-06 & 577.0 & 293.7 $\pm$ 0.3 & 3.90$\pm$0.09 & 15.63$\pm$0.17 & F814W \\
 & 2000-05-06 & 570.0 & 293.7 $\pm$ 0.3 & 3.54$\pm$0.09& 13.63$\pm$0.17 & F1042M\\

 &  & & &   \\

SDSS2335583-001304      & 2001-06-29 & 56.8 & 8.1 $\pm$ 1.2 & 1.00$\pm$0.11 & 19.18$\pm$0.18 & F814W \\

 &  & & &   \\
\enddata
\tablenotetext{a}{The object was observed in a corner of the HST/WF, where the pixel scale is coarser and the distortions higher, hence implying higher uncertainties, explaining the discrepancy between the values obtained in the two filters}
\tablenotetext{b}{The secondary fell on a boundary between two pixels and was very elongated on both images, thus increasing the uncertainties and explaining the discrepancy between the values obtained in the two filters. As the image in the F814W were more sensitive and as the secondary is brighter in that band, we hereafter keep the corresponding value as final result}
\tablenotetext{c}{A cosmic ray event fall too close to the primary to allow precise measurements}
\tablenotetext{d}{1$\sigma$ uncertainties: refer to the description given in section \ref{dataanalysis}, unless a different value is specified in the table.}
\tablenotetext{e}{The uncertainties on these values includes both the uncertainties on the measure of the magnitude of the whole system and the uncertainties on the difference of magnitude.}
\end{deluxetable}
\end{center}

\clearpage

\begin{center}
\begin{deluxetable}{lcc}
\tabletypesize{\scriptsize}
\tablecaption{Photometry of the unresolved objects of program GO8720 \tablenotemark{a}\label{photom_go8720}}
\tablewidth{0pt}
\tablehead{
\colhead{Name} & \colhead{$m_{F814W}$ \tablenotemark{b}}   & \colhead{$m_{F675W}$ \tablenotemark{b}}
}
\startdata
DENISPJ0243-5432        &       17.1    &       19.5    \\

DENISPJ0301-5903        &       16.6    &       19.2    \\
DENISPJ0314-4623        &       17.9    &       \nodata \\
DENISPJ0426-5735        &       18.3    &       19.9    \\
DENISPJ0441-0211        &       18.7    &       20.7    \\
DENISPJ1016-2714        &       18.0    &       20.4    \\
DENISPJ1047-1815        &       17.4    &       19.8    \\
DENISPJ1048-3956        &       12.4    &       15.1    \\
2MASSWJ1145+2317        &       18.6    &       20.6    \\
DENISPJ1154-3400        &       17.6    &       19.8    \\
DENISPJ1216-1257        &       18.0    &       20.0    \\
DENISPJ1315-2513        &       18.1    &       20.7    \\
2MASSWJ1342+1751        &       19.2    &       21.6    \\
2MASSIJ1346-0031        &       20.8    &       22.9    \\
DENISPJ1412-0433        &       18.2    &       20.8    \\
2MASSWJ1439+1826        &       19.4    &       \nodata \\
SDSSJ1440+0021          &       19.2    &       21.5    \\
DENISPJ1619+0050        &       17.6    &       20.0    \\
DENISPJ1919-4134        &       20.0    &       \nodata \\
DENISPJ2023-1815        &       18.0    &       19.5    \\
DENISPJ2215-0809        &       17.9    &       20.0    \\
DENISPJ2329-5408        &       18.3    &       20.4    \\
\enddata
\tablenotetext{a}{Not all object were bright enough to do photometry.}
\tablenotetext{b}{Uncertainties on these values are $\sim$0.1 mag}
\end{deluxetable}
\end{center}

\clearpage

\begin{center}
\begin{deluxetable}{lcc}
\tabletypesize{\scriptsize}
\tablecaption{Photometry of the unresolved objects of program GO8581\label{photom_go8581}\tablenotemark{a}}
\tablewidth{0pt}
\tablehead{
\colhead{Name} & \colhead{$m_{F814W}$ \tablenotemark{b}}   & \colhead{$m_{F1042M}$ \tablenotemark{b}}
}
\startdata
SDSS0019+0030  &  18.1  &  16.0 \\
SDSS0344+0111  &  17.8  &  14.6 \\
SDSS1043+0001  &  19.2  &  \nodata \\
SDSS1435-0043  &  19.7  &  \nodata \\
SDSS1435-0046  &  19.1  &  \nodata \\
SDSS1548-0029  &  18.2  &  16.6 \\
SDSS1653+6231  &  17.7  &  15.3 \\
SDSS1723+6406  &  18.9  &  \nodata \\
SDSS0330-0025  &  18.8  &  16.4 \\
SDSS0539-0059  &  17.5  &  15.0 \\
SDSS1515-0030  &  17.0  &  15.0 \\
2MASS0326-2102  &  20.0  &  \nodata \\
2MASSW0010+3436  &  18.1  &  16.3 \\
2MASSW0028+1501  &  20.1  &  17.5 \\
2MASSW0030-1450  &  18.3  &  \nodata \\
2MASSW0033-1521  &  18.0  &  \nodata \\
2MASSW0208+2542  &  17.2  &  15.0 \\
2MASSW0224+2537  &  19.7  &  \nodata \\
2MASSW0328+2302  &  20.4  &  16.5 \\
2MASSW0335+2342  &  14.7  &  12.8 \\
2MASSW0337-1758  &  19.4  &  16.3 \\
2MASSW0345+2540  &  17.2  &  14.8 \\
2MASSW0350+1818  &  15.6  &  13.6 \\
2MASSW0355+2257  &  19.6  &  \nodata \\
2MASSW0753+2917  &  18.8  &  16.6 \\
2MASSW0801+4628  &  19.6  &  16.4 \\
2MASSW0829+2655  &  20.8  &  18.2 \\
2MASSW0832-0128  &  17.4  &  14.9 \\
2MASSW0914+2238  &  18.4  &  16.0 \\
2MASSW0951+3558  &  21.0  &  \nodata \\
2MASSW1102-2359  &  20.5  &  \nodata \\
2MASSW1104+1959  &  17.9  &  15.8 \\
2MASSW1108+6830  &  16.6  &  \nodata \\
2MASSW1239+2029  &  17.3  &  15.2 \\
2MASSW1403+3007  &  15.4  &  13.4 \\
2MASSW1411+3936  &  17.8  &  15.5 \\
2MASSW1412+1633  &  17.0  &  14.9 \\
2MASSW1434+1940  &  18.1  &  15.1 \\
2MASSW1438-1309  &  19.0  &  17.0 \\
2MASSW1438+6408  &  16.0  &  13.8 \\
2MASSW1457+4517  &  16.0  &  13.9 \\
2MASSW1506+1321  &  16.8  &  14.5 \\
2MASSW1526+2043  &  18.9  &  16.5 \\
2MASSW1550+3041  &  15.4  &  13.3 \\
2MASSW1551+6457  &  15.8  &  13.6 \\
2MASSW1627+8105  &  16.0  &  13.7 \\
2MASSW1635+4223  &  15.7  &  12.2 \\
2MASSW1656+2835  &  20.5  &  \nodata \\
2MASSW1707+4301  &  17.1  &  14.9 \\
2MASSW1707+6439  &  15.4  &  13.3 \\
2MASSW1710+2107  &  18.4  &  15.8 \\
2MASSW1711+2232  &  20.5  &  \nodata \\
2MASSW1743+2127  &  19.2  &  \nodata \\
2MASSW1743+5844  &  17.2  &  14.1 \\
2MASSW1841+3117  &  19.5  &  16.8 \\
2MASSW2054+1519  &  19.6  &  \nodata \\
2MASSW2057+1715  &  19.3  &  17.1 \\
2MASSW2158-1550  &  18.5  &  14.9 \\
2MASSW2206-4217  &  19.1  &  15.8 \\
2MASSW2208+2921  &  19.4  &  \nodata \\
2MASSW2234+2359  &  16.3  &  14.1 \\
2MASSW2242+2542  &  18.0  &  15.2 \\
2MASSW2244+2043  &  20.2  &  \nodata \\
2MASSW2306-0502  &  14.0  &  12.0 \\
2MASSW2309+1549  &  17.8  &  15.8 \\
2MASSW2349+1224  &  15.3  &  13.2 \\

\enddata
\tablenotetext{a}{Not all object were bright enough to do photometry.}
\tablenotetext{b}{Uncertainties on these values are $\sim$0.1 mag}
\end{deluxetable}
\end{center}

\clearpage

\begin{center}
\begin{deluxetable}{lcc}
\tabletypesize{\scriptsize}
\tablecaption{Photometry of the unresolved objects of program GO8146\label{photom_go8146}}
\tablewidth{0pt}
\tablehead{
\colhead{Name} & \colhead{$m_{F814W}$ \tablenotemark{a}}
}
\startdata
2MASSW0036+1821 & 15.9 \\
2MASSW0708+2950 & 20.4 \\
2MASSW0740+3212 & 19.7 \\
2MASSW0820+4500 & 20.1 \\
2MASSW0825+2115 & 18.8 \\
2MASSW0913+1841 & 19.3 \\
2MASSW0928-1603 & 18.7 \\
2MASSW1123+4122 & 19.6 \\
2MASSW1155+2307 & 19.5 \\
2MASSW1328+2114 & 19.9 \\
2MASSW1338+4140 & 17.5 \\
2MASSW1343+3945 & 19.9 \\
2MASSW1439+1929 & 16.0 \\
2MASSW1507-1627 & 16.4 \\
2MASSW1632+1904 & 19.7 \\
2MASSW1726+1538 & 19.3 \\

\enddata
\tablenotetext{a}{Uncertainties on these values are $\sim$0.1 mag}
\end{deluxetable}
\end{center}

\clearpage

\begin{center}
\begin{deluxetable}{l c c c c}
\tabletypesize{\scriptsize}
\tablecaption{Resolved Field Ultra-cool Dwarfs Binaries \label{table_periods}}
\tablewidth{0pt}
\tablehead{
\colhead{Name} & \colhead{Distance}   & \colhead{Sep.} & Semi-major & \colhead{Period\tablenotemark{d}} \\
 & (pc) & (A.U) & axis\tablenotemark{c}   (A.U) & (yrs) \\
}
\startdata
DENISPJ0205-1159 & 19.8 & 7.3 & 9.2 & 74.6  \\
DENISPJ0357-4417 & 22.2 \tablenotemark{b} & 2.2 & 2.8 & 10.5    \\
2MASSW0746+2000 & 12.3 & 2.7 & 3.4 & 14.0   \\
2MASSW0850+1057 & 41.0 & 6.4 & 8.1 & 61.6  \\
2MASSW0856+2235 & 34.7 \tablenotemark{b} & 3.5 & 4.4 & 20.8  \\
2MASSW0920+3517 & 20.1 \tablenotemark{b} & 1.5 & 1.9 & 7.2   \\
DENISPJ1004-1146 & 46.8\tablenotemark{b} & 6.8 & 8.6 & 56.5  \\
2MASSW1017+1308 & 21.4 \tablenotemark{b} & 2.3 & 2.9 & 11.0 \\
2MASSW1127+7411 & 14.6 \tablenotemark{b} & 3.8 & 4.8 & 23.5 \\
2MASSW1146+2230 & 27.2 & 8.0 & 10.1 & 92.6   \\
DENISPS1228-1547& 20.2 & 5.1 & 6.6 & 49.0  \\
2MASSW1239+5515 & 21.3 \tablenotemark{b} & 4.7 & 5.9 & 32.0 \\ 
2MASSW1311+8032 & 13.7 \tablenotemark{b} & 4.4 & 5.5 & 28.7 \\
2MASSW1426+1557 & 26.7 \tablenotemark{b} & 4.3 & 5.4 & 33.3   \\
2MASSW1430+2915 & 29.4 \tablenotemark{b} & 2.7 & 3.4 & 13.9 \\
DENISPJ1441-0945 & 29.2\tablenotemark{b} & 11.2 & 14.1 & 118.0  \\
2MASSW1449+2355 & 63.7 \tablenotemark{b} & 8.7 & 11.0 & 81.1 \\
2MASSW1600+1708 & 60.6 \tablenotemark{b} & 3.4 & 4.3 & 19.9 \\
2MASSW1728+3948 & 20.4 \tablenotemark{b} & 2.8 & 3.4 & 13.8 \\
2MASSW2101+1756 & 23.2 \tablenotemark{b} & 5.6 & 7.1 & 41.9 \\
2MASSW2140+1625 & 12.9 \tablenotemark{b} & 2.1 & 2.6 & 10.5 \\
2MASSW2147+1431 & 21.8 \tablenotemark{b} & 7.2 & 9.1 & 61.7 \\
2MASSW2206-2047 & 22.2 \tablenotemark{b} & 3.6 & 4.5 & 22.6  \\
2MASSW2331-0406 & 26.2 & 15.1 & 19.0 & 211.0 \\
SDSS2335-0013   & 20 \tablenotemark{a}   & 1.1 & 1.4 & 3.7  \\
CFHT-PL-18      & 105 \tablenotemark{b} & 35.0 & 44.1 & 641 \\

\enddata
\tablenotetext{a}{The distance cannot be estimated by any method. We assume an average distance of 20 pc.}
\tablenotetext{b}{Photometric distances evaluated as explained in section \ref{spt}.}
\tablenotetext{c}{Semi-major axis calculated by multiplying the projected separation by the correction factor 1.26 as explained in \citet{fischer92}, in order to take in account inclination, orbital angle, etc.}
\tablenotetext{d}{This period is just an estimate, calculated as explained in the text.}
\end{deluxetable}
\end{center}

\clearpage

\begin{center}
\begin{deluxetable}{l l c c c c c}
\tabletypesize{\scriptsize}
\tablecaption{Ultra-cool Dwarfs Binaries associated to another star \label{triple}}
\tablewidth{0pt}
\tablehead{
\colhead{Name} & \colhead{Name of the} & \colhead{Distance\tablenotemark{a}}   & \colhead{Sep. (A/B)} & \colhead{Sep. (B/C)} & \colhead{Period (B/C)\tablenotemark{b}} \\
 & primary & (pc) & (A.U) & (A.U) & (yrs) \\
}
\startdata
2MASSW1112+3548 & Gl 417 & 21.7 & 1953 & 1.5 & 5.8 \\
Gliese 569B     & Gliese 569A & 9.8  & 49 & 1.1 & 2.9 \\
HD 130948BC     & HD 130948 & 17.9 & 47.2 & 2.4 & 10 \\
\enddata
\tablenotetext{a}{Hipparcos trigonometric parallax of the Primary \citep{perryman97}}
\tablenotetext{b}{This period is just an estimate, calculated as explained in the text.}

\end{deluxetable}
\end{center}

\clearpage
\begin{center}
\tablecaption{\label{table_d1228_d0205}}
\begin{deluxetable}{l l l l l}
\tablecolumns{5}
\tabletypesize{\scriptsize}
\tablewidth{0pt}
\tablehead{
\multicolumn{4}{c}{DENISPJ122815.4-154730} & \colhead{Ref.}
}
\startdata
Separation (mas) : & P.A ($^{\circ}$) : & Date : & & \\
 $ 275 \pm 2$ &  $41.0 \pm 0.2$ & 02-06-1998 &  & (1) \\
 $246 \pm 20$  \tablenotemark{a} &  $23 \pm 2$  \tablenotemark{a} &  03-04-2001 & & (4) \\
 $255.4 \pm 2.8$ & $18.3 \pm 0.3$ & 16-06-2001 & & (4) \\
 & & & & \\
Distance (pc) :    & $18 \pm 4 $  & & & (1) \\
                 & $20.2 \pm 1$ & & &   (2) \\
 & & & & \\
Proper Motion (mas/yr) :   & $\mu=224 \pm 1.3$ & &  & (2) \\
 &                         $P.A=143.3 \pm 0.3$  &  & & (2) \\
 & & & & \\
Parallax (mas) : & $49.4 \pm 1.9$ & & & (2) \\
 & & & &  \\
$v.sin(i)$ $(km.s^{-1})$ :       & $22 \pm 2.5$ & & & (3) \\
 & & & &  \\

\cutinhead{DENISPJ020529-115930}

Distance (pc) :    & $19.8 \pm 0.6$ & & &       (2) \\
 & & & & \\
Proper Motion (mas/yr) :  &  $\mu=437.8 \pm 0.8$  &  & & (2) \\
 &                         $P.A=82.8 \pm 0.2$   & &  & (2) \\
 & & & & \\
Parallax (mas) : & $59.4 \pm 2.6$ & & & (2) \\
 & & & &  \\
$v.sin(i)$ $(km.s^{-1})$ :       & $22 \pm 5$ & & & (3) \\
 & & & &  \\
$T_{eff}$ : & $1700 \sim 1800$ K & & & (3)\\

\enddata
\tablerefs{
(1) \citet{martin99a}, (2) \citet{dahn2002}, (3) \citet{basri2000}, (4) this paper.}
\tablenotetext{a}{The object was observed in a corner of the HST/WF, where the pixel scale is coarser and the distortions higher, hence implying higher uncertainties}

\end{deluxetable}
\end{center}

\clearpage

\begin{deluxetable}{lcccccccc}
\tablecolumns{7}
\tablewidth{0pc}
\tablecaption{Examples of mass ratios based on the DUSTY model \label{model}}
\tablehead{
\colhead{$\Delta m_{F814W}$} & \colhead{}& \colhead{1.5} &   \colhead{}   & \colhead{3.0} & \colhead{} & \colhead{4.0} & \colhead{} & \colhead{5.5}\\
\cline{1-9}
\colhead{Age} & \colhead{}  & \colhead{M$_{2}/$M$_{1}$} & \colhead{} & \colhead{M$_{2}/$M$_{1}$} & \colhead{} & \colhead{M$_{2}/$M$_{1}$} & \colhead{} & \colhead{M$_{2}/$M$_{1}$}
}
\startdata
0.5 Gyr & & 75$\%$ & & 62$\%$ & & 55$\%$ & & 45$\%$\\
1.0 Gyr & & 85$\%$ & & 73$\%$ & & 68$\%$ & & 61$\%$\\
5.0 Gyrs & & 95$\%$ & & 77$\%$ & & 74$\%$ & & 71$\%$\\
\enddata
\tablecomments{Examples of mass ratios based on the DUSTY model convolved with the HST/WFPC F814W filter, corresponding to luminosity ratios of $\Delta m_{F814W} =$ 1.5, 3.0, 4.0, and 5.5 mag, for a primary of 0.1 M$_{\sun}$, according to the DUSTY model of \citet{chabrier00}}
\end{deluxetable}

\end{document}